\documentclass[a4paper,fleqn]{cas-dc}

\usepackage{xcolor}
\usepackage[utf8]{inputenc}
\usepackage[english]{babel}
\usepackage{float}
\usepackage{caption}
\usepackage{subcaption}

\usepackage{tikz}
\usepackage{stfloats}

\usepackage{amsthm}
\newtheorem{theorem}{Theorem}[section]
\newproof{pf}{Proof}
\newproof{pot}{Proof of Theorem}

\usepackage{amsmath,amssymb,amsfonts}
\usepackage{amsmath}
\usepackage{textcomp}

\usepackage{amsthm}
\usepackage[square, sort, numbers]{natbib}
\usepackage{algorithm}
\usepackage[noend]{algpseudocode}

\makeatletter
\def\BState{\State\hskip-\ALG@thistlm}
\makeatother

\algnewcommand\algorithmicswitch{\textbf{switch}}
\algnewcommand\algorithmiccase{\textbf{case}}
\algnewcommand\algorithmicassert{\texttt{assert}}
\algnewcommand\Assert[1]{\State \algorithmicassert(#1)}%
\algdef{SE}[SWITCH]{Switch}{EndSwitch}[1]{\algorithmicswitch\ #1\ \algorithmicdo}{\algorithmicend\ \algorithmicswitch}%
\algdef{SE}[CASE]{Case}{EndCase}[1]{\algorithmiccase\ #1}{\algorithmicend\ \algorithmiccase}%
\algtext*{EndSwitch}%
\algtext*{EndCase}%

 \usepackage{amssymb}
\usepackage{amsmath}

\usepackage{tabularx}

\usepackage[utf8]{inputenc}
\usepackage{pifont}
\newcommand{\cmark}{\ding{51}}%
\newcommand{\xmark}{\ding{55}}%

\usepackage{enumitem}
\usepackage{comment}

\def\tsc#1{\csdef{#1}{\textsc{\lowercase{#1}}\xspace}}
\tsc{WGM}
\tsc{QE}
\tsc{EP}
\tsc{PMS}
\tsc{BEC}
\tsc{DE}
\begin{document}
\let\WriteBookmarks\relax
\def\floatpagepagefraction{1}
\def\textpagefraction{.001}

\shorttitle{Communication-Enabled DRL to Optimise EE in UAV-Assisted Networks}

\shortauthors{Omoniwa B., Galkin B. \& Dusparic I.}

\title [mode = title]{Communication-Enabled Deep Reinforcement Learning to Optimise Energy-Efficiency in UAV-Assisted Networks}                      
%%\tnotemark[1,2]

\tnotetext[1]{This work was supported, in part, by the Science Foundation Ireland (SFI) Grants No. 16/SP/3804 (Enable) and 13/RC/2077\_P2 (CONNECT Phase 2), the National Natural Science Foundation Of China (NSFC) under the SFI-NSFC Partnership Programme Grant Number 17/NSFC/5224. Dr. Galkin's work was funded by SFI under the MISTRAL project, grant number 21/FIP/DO/9949, as well as the project "GUARD: Drug Interdiction Using Smart Drones" funded under Enterprise Ireland's Disruptive Technologies Innovation Fund, Project ref. DT20200268A.}

%\tnotetext[2]{The second title footnote which is a longer text matter  to fill through the whole text width and overflow into   another line in the footnotes area of the first page.}


\author[1]{Babatunji Omoniwa}[orcid=0000-0003-3508-3689]

\cormark[1]


\ead{omoniwab@tcd.ie}

%\ead[url]{omoniwab@tcd.ie}

\credit{Conceptualization of this study, Methodology, Software, Writing - Original draft preparation}

\affiliation[1]{organization={Trinity College Dublin},
    %addressline={Radarweg 29}, 
    city={Dublin},
    % citysep={}, % Uncomment if no comma needed between city and postcode
    %postcode={1043 NX}, 
    country={Ireland}}

\author[2]{Boris Galkin}[orcid=0000-0002-6755-7781]
\ead{boris.galkin@tyndall.ie}
\credit{Methodology, Writing - Original draft preparation}

\author[1]{Ivana Dusparic}[orcid=0000-0003-0621-5400]

\ead{ivana.dusparic@scss.tcd.ie}
%\ead[URL]{www.sayahna.org}

\credit{Conceptualization of this study, Methodology}

\affiliation[2]{organization={Tyndall National Institute},
    city={Cork},
    % citysep={}, % Uncomment if no comma needed between city and postcode
    %postcode={695014}, 
    %state={Trivandrum},
    country={Ireland}}

\cortext[cor1]{Corresponding author}
%\cortext[cor2]{Principal corresponding author}

%\fntext[fn1]{This is the first author footnote. but is common to third author as well.}
%\fntext[fn2]{Another author footnote, this is a very long footnote and   it should be a really long footnote. But this footnote is not yet  sufficiently long enough to make two lines of footnote text.}

%\nonumnote{This note has no numbers. In this work we demonstrate $a_b$  the formation Y\_1 of a new type of polariton on the interface   between a cuprous oxide slab and a polystyrene micro-sphere placed  on the slab.  }

\begin{abstract}
Unmanned aerial vehicles (UAVs) are increasingly deployed to provide wireless connectivity to static and mobile ground users in situations of increased network demand or points of failure in existing terrestrial cellular infrastructure. However, UAVs are energy-constrained and experience the challenge of interference from nearby UAV cells sharing the same frequency spectrum, thereby impacting the system's energy efficiency (EE).
%%gaps
Recent approaches focus on optimising the system's EE by optimising the trajectory of UAVs serving only static ground users and neglecting mobile users. Several others neglect the impact of interference from nearby UAV cells, assuming an interference-free network environment. Furthermore, some works assume global spatial knowledge of ground users' location via a central controller (CC) that periodically scans the network perimeter and provides real-time updates to the UAVs for decision-making. However, this assumption may be unsuitable in disaster scenarios since it requires significant information exchange between the UAVs and CC. Moreover, it may not be possible to track users' locations in a disaster scenario. Despite growing research interest in decentralised control over centralised UAVs' control, direct collaboration among UAVs to improve coordination while optimising the systems' EE has not been adequately explored.
%%%Our proposed approach
To address this, we propose a direct collaborative communication-enabled multi-agent decentralised double deep Q-network (CMAD--DDQN) approach. The CMAD--DDQN is a collaborative algorithm that allows UAVs to explicitly share their telemetry via existing 3GPP guidelines by communicating with their nearest neighbours. This allows the agent-controlled UAVs to optimise their 3D flight trajectories by filling up knowledge gaps and converging to optimal policies. We account for the mobility of ground users, the UAVs' limited energy budget and interference in the environment. Our approach can maximise the system's EE without hampering performance gains in the network. Simulation results show that the proposed approach outperforms existing baselines in terms of maximising the systems' EE  without degrading coverage performance in the network. The CMAD--DDQN approach outperforms the MAD-DDQN that neglects direct collaboration among UAVs, the multi-agent deep deterministic policy gradient (MADDPG) and random policy approaches that consider a 2D UAV deployment design while neglecting interference from nearby UAV cells by about 15\%, 65\% and 85\%, respectively.
\end{abstract}


\maketitle

\section{Introduction}
It is envisaged that machine-to-machine (M2M) connections will grow 2.4-fold, from 6.1 billion in 2018 to 14.7 billion by 2023~\cite{5GPP_AI_ML_networks2020}. Unmanned aerial vehicles (UAVs) can play a vital role in supporting the Internet-of-Things (IoT) networks by providing ubiquitous connectivity to static and mobile ground devices~\cite{Omoniwa2018}.
For instance, the deployment of UAVs to provide wireless connectivity to ground users is gaining significant research attention~\cite{Omoniwa2022_Letters} -- \cite{Cui2019MARLtransactionUAV}. UAVs' deployment can complement cellular networks by accommodating the projected growth of connected things. Specifically, UAVs' adjustable altitude and mobility make them suitable candidates for flexible deployment as aerial base stations in the event of increased network demand, points-of-failure in existing terrestrial infrastructure, or emergencies~\cite{Omoniwa_DQLCI_2021, Jin_2023}.~For example, UAVs may be deployed to provide coverage to users in post-disaster scenarios where existing terrestrial infrastructures are damaged~\cite{Tang_2020}.~However, it is challenging to conserve the energy of UAVs during prolonged coverage tasks, considering their limited onboard battery capacity. UAVs may deplete energy during propulsion for flying and hovering, and during communication~\cite{Mozaffari2017UAV}. To derive the full benefit of UAV deployments, recent research has focused on addressing some main challenges, they include the 3D trajectory optimisation~\cite{Wang2021_MADDPG, Liu2019UAVqlearning, Oubbati_2023}, energy efficiency (EE) optimisation~\cite{Omoniwa2022_Letters, Liu2020UAVdistributed}, and coverage optimisation~\cite{Omoniwa_DQLCI_2021, Liu2020UAVdistributed, Oubbati_2023}. As energy-constrained UAVs fly in the sky, they may encounter interference from nearby UAV cells or other access points sharing the same frequency band, thereby affecting the system's EE~\cite{Galkin2020UAVDeepLearning}. There has been significant research on optimising the EE in UAV-assisted networks. However, many of these works neglect the impact of interference on the system's performance.

Compared with a terrestrial cellular communication network, channel modelling for an airborne, UAV-assisted
wireless system is more challenging due the mobility and direct line-of-sight (LoS) communication link from nearby UAVs~\cite{Zhang2021UAV}. Furthermore, the adoption of UAVs for communication may require jointly finding the optimal 3D deployment plan, energy  and interference management strategy~\cite{Bouzid_2023}. Crucially, UAVs require robust strategies to provide ubiquitous wireless coverage to static and mobile ground users in this dynamic environment. Unlike previous work that assumes global spatial knowledge of ground users' location through a central controller that periodically scans the network perimeter and provides real-time updates to the UAVs for decision-making, we focus on a decentralised approach suitable in emergency scenarios where there may be service outage due to failure in the controller, or loss of UAVs' control packets due traffic congestion in the network. Moreover, in such scenarios, it is difficult to keep track of the location of all ground users in real time. To simplify the model, recent approaches that optimise the system's EE consider a 2D trajectory optimisation design of UAVs serving static users in an interference-free network environment. This may be based on the assumption that each operating UAV is assigned a unique frequency band. However, this assumption is impractical as radio spectrum is a scarce resource. Hence, we assume that UAVs serving as aerial base stations may have to share same frequency band. However, this introduces the challenge of interference in the shared network environment. Nevertheless, UAVs require robust strategies to optimise their flight trajectory while providing coverage to ground users in a dynamic environment. Multi-Agent Reinforcement Learning (MARL) has been shown to perform well in decision-making tasks in such a dynamic environment~\cite{Omoniwa2022_Letters, Liu2020UAVdistributed, Liu2019UAVqlearning}. To improve the performance of the decentralised control, several methods have been studied~\cite{Ming_MARL_1993}. In this work, we adopt a MARL approach and propose a direct collaborative communication-enabled multi-agent decentralised double deep Q-network (CMAD--DDQN) algorithm to maximise the system's EE by optimising the 3D trajectory of each UAV, the energy consumed and the number of connected static and mobile ground users over a series of time-steps, while taking into account the impact of interference from nearby UAV cells.

%%%%%%%%%%%%%%%%%%Contribution
In our previous work~\cite{Omoniwa_DQLCI_2021}, we considered a decentralised MARL where there was no direct collaboration among UAVs and other agents are treated as a part of the environment, with the reward of each agent reflecting the coverage performance in its neighbourhood. However, the approach~\cite{Omoniwa_DQLCI_2021} ignores the potential benefit of direct collaboration among agents. Moreover, finding a globally optimal solution for agents with partial information is known to be intractable~\cite{Yagan2007}. As an extension to our prior work~\cite{Omoniwa2022_Letters}, we leverage agents' capability to communicate with neighbours to maximise the system's EE by jointly optimising the number of connected ground users and the energy consumption in the network. The incorporation of collaborative algorithms into MARL can allow the agents to assist each other in filling the knowledge gaps by
exchanging information that could improve the decision-making of UAVs over a series of time-steps~\cite{Zhang2021_survey}. However, several real-time applications place considerable restrictions on communication, especially in terms of both throughput and latency. Nevertheless, communication has extensively been used to address the non-stationarity issue in the multi-agent learning process~\cite{Szer2004_Communication}.

Multi-agent learning is challenging in itself, requiring agents to learn their policies while taking into account the consequences of the actions of others. The authors in~\cite{Liu2020UAVdistributed, Wang2021_MADDPG, Liu2018UAV} proposed a multi-agent deep deterministic policy gradient (MADDPG) approach to improve the system's EE as UAVs hover at fixed altitudes while providing coverage to static ground users in an interference-free network environment. This problem becomes even more challenging in an interference-limited network environment, where interference from nearby UAV cells impacts the system's EE. Hence, we propose a direct collaborative communication-enabled multi-agent decentralised double deep Q-network (CMAD--DDQN) approach where each agent relies on its local observations, as well as the information it receives from its nearby UAVs for decision-making. The communicated information from the nearby UAVs will contain the number of connected ground users, instantaneous energy value, and distances from nearby UAVs in each time step. We propose an approach where each agent executes actions based on state information. We assume a two-way communication link among neighbouring UAVs~\cite{3GPPstandard_uav_comm}. Although the 3GPP system provides a methodology to set up and optimise neighbour relations with little or no human intervention~\cite{3GPPstandard_neighborRelations} and to allow a 3rd party to request and obtain real-time monitored status information (e.g., position, communication link status, power consumption) of a UAV~\cite{3GPPstandard_uav_comm}, to the best of our knowledge this work is first to investigate the impact of collaborations on the system's EE using the communication mechanism based on the existing 3GPP standard~\cite{3GPPstandard_neighborRelations}. This paper has three main contributions given as follows:
\begin{itemize}[leftmargin=*]
  \item We propose a direct collaborative CMAD--DDQN approach that relies on local observations from each UAV and the explicitly-communicated information from its neighbours for decision-making. We adopt a collaborative algorithm based on an existing 3GPP standard~\cite{3GPPstandard_neighborRelations} that allows agents to collaborate by exchanging information with their nearest neighbours to improve the system's EE by jointly optimising each UAV's 3D trajectory, the number of connected ground users, and the energy consumed by the UAVs in a shared dynamic environment. 
  \item We consider a realistic model of the agent's environment, taking into consideration the dynamic and interference-limited nature of the wireless environment. Unlike in previous work that consider the deployment of static users~\cite{Liu2020UAVdistributed} or fully synthetic ground users' distribution~\cite{Omoniwa2022_Letters}, we consider a real-world deployment of static and mobile end-users in an area of Dublin, Ireland. Furthermore, we leverage widely used mobility models (the random walk (RW), random waypoint (RWP) and the Gauss–Markov (GMM) mobility models)~\cite{Song2010, Camp2002} to depict pedestrian movements. 
  \item We evaluated the proposed CMAD--DDQN approach by comparing it with the MAD--DDQN~\cite{Omoniwa2022_Letters} that ignores direct collaboration among UAVs, the MADDPG~\cite{Liu2020UAVdistributed} that considers a 2D UAV deployment design while neglecting interference from nearby UAV cells, and the random policy. Results show that our proposed approach can significantly improve the total system's EE while jointly optimising the 3D trajectory, number of connected users, and the energy consumed by the UAVs serving ground users under a strict energy budget.
\end{itemize}
The remainder of this work is organised as follows.~In Section II, we present related work. The environment model is provided in Section III. We discuss the proposed decentralised MARL approach for EE optimisation in Section IV. In Section V, we present the simulation setup and evaluation plan. We discuss and analyse the results in Section VI. Section VII concludes the paper and outlines future directions.
 
\section{Related Work}
\label{subsec:Relatedwork}
Energy efficiency (EE) optimisation in UAV-assisted networks has been studied recently~\cite{Liu2020UAVdistributed, Liu2018UAV, Omoniwa2022_Letters}. The works~\cite{Zeng2019UAV, Yao2019UAV, Hua2019UAV} proposed classical optimisation techniques to optimise the EE of a single UAV deployed to provide wireless service to static ground users. A similar technique was used in a relay scenario~\cite{Sun2021UAV} to jointly optimise the energy and trajectory of UAVs while transferring information
from source ground users to corresponding destination ground users via the UAV relay. In \cite{Jing2021UAV}, an iterative algorithm was proposed to optimise the trajectory of a fixed-wing UAV base station deployed at fixed altitudes while optimising the transmit power in each iteration. However, these single UAV models may not be applicable in larger geographical areas where multiple UAVs are deployed to serve ground users. 
Moreover, it is anticipated that future UAV networks will have multiple UAVs flying in the sky, and possibly sharing the same frequency spectrum. Table \ref{tab:MARLUAV} shows a summary of related work on multi-UAV deployments.

Recent research focuses on optimising EE in multi-UAV networks~\cite{Liu2020UAVdistributed, Mozaffari2017UAV, Liu2018UAV}. An iterative algorithm was proposed in~\cite{Mozaffari2017UAV} to minimise the energy consumption of UAV base stations providing coverage to static ground users. Game theory was proposed in~\cite{Ruan2018UAV} to optimise the system's EE while maximising the ground area covered by the UAVs irrespective of the presence of ground users. However, this work may only be suitable in scenarios with an unlimited energy budget or cost of UAVs deployment. Furthermore, these works rely on a ground controller that supports the decision-making of the UAVs, hence making emergency deployment impractical due to the significant amount of information exchanged between the UAVs and the controller. Moreover, tracking ground user locations at each time step may be difficult. In \cite{Luo2021UAV}, a classical optimisation method was used to minimise the energy consumption of static ground users by optimizing the UAVs' trajectory. As energy-constrained UAVs fly in the sky, they may encounter interference from nearby UAV cells or other access points sharing the same frequency band, thereby affecting the system's EE~\cite{Galkin2020UAVDeepLearning}.

The adoption of machine learning to solve complex multi-UAV deployment problems is gaining research attention~\cite{Cui2019MARLtransactionUAV}. Specifically, multi-agent reinforcement learning (MARL) approaches have been used in several works to optimise the system's EE.
The work~\cite{Hu2020_Cooperative_UAVs} proposed a
distributed sense-and-send protocol, where the UAVs determine
their trajectories by selecting from a discrete set of tasks and a continuous set of locations for sensing and transmission. However, the UAVs relied on a feedback mechanism from the central base station to execute the next task, thus leading to significant communication overhead in the centralised architecture. Furthermore, the authors did not consider the impact of interference from neighbouring UAVs in this shared multi-agent environment. Our prior work applied a distributed Q-learning approach~\cite{Omoniwa_DQLCI_2021} to optimise the energy utilisation of UAVs providing coverage to ground users without taking into account the system's EE. The work~\cite{Omoniwa_DQLCI_2021} considered a limited number of deployed independent learning agent-controlled UAVs that have no mechanism for direct collaboration. These UAVs were deployed to serve mobile ground users that follow the RWP model. To solve this problem, a deep reinforcement learning (DRL) approach~\cite{Galkin2020UAVDeepLearning} was proposed in our recent work for intelligent UAV cellular users to base station association, allowing a UAV flying over an urban area to intelligently connect to underlying base stations. A deep meta reinforcement learning-based offloading algorithm was proposed in \cite{Qu_UAV_2022} to make fine-grained offloading decisions in a dynamic environment. In our prior work~\cite{Galkin2022fixedWingUAV}, a DRL-based approach was proposed to optimise the EE of fixed-winged UAVs that move in circular orbits and are typically not able to hover like the rotary-winged UAVs. Moreover, the attention was on UAVs deployed to provide connectivity to static ground users. The distributed DRL work in~\cite{Liu2020UAVdistributed} improved on the centralised approach in~\cite{Liu2018UAV}, where all deployed UAVs are controlled by a single autonomous agent. The authors in~\cite{Liu2020UAVdistributed, Wang2021_MADDPG, Liu2018UAV} proposed a deep deterministic policy gradient (DDPG) approach to optimise the system's EE as the UAVs hover at fixed altitudes while serving static ground users in an interference-free network environment. Although the approaches \cite{Liu2020UAVdistributed, Wang2021_MADDPG, Liu2018UAV} improve the coverage performance of UAVs, they focus on the 2D trajectory optimisation of the UAVs serving static ground users. The authors in~\cite{Oubbati_2023} also proposed a DDPG-based solution to optimise the flight trajectory and coverage performance of UAVs serving mobile users, with the mobility of the ground users modelled to follow the RWP and reference point group mobility (RPGM) models.

\begin{table*}[t]
    \centering 
    \footnotesize
    \caption{Related Work on Multiple UAVs Deployed as Aerial Base Stations.}
    \begin{tabular}{lllllll} \hline
        \textit{\textbf{Paper}} & \textit{\textbf{Approach}} & \textit{\textbf{Training}} & \textit{\textbf{Execution}} & \textit{\textbf{Control}} & \textit{\textbf{Flight}} & \textit{\textbf{User}}\\ 
        &  &  &  & \textit{\textbf{Overhead}} & \textit{\textbf{Trajectory}} & \textit{\textbf{Data}}\\
        \hline \hline
         Liu C. H. \textit{et al.} \cite{Liu2018UAV}  & DDPG & Centralised & Centralised & Global & 2D & Synthetic\\ 
         Liu C. H. \textit{et al.} \cite{Liu2020UAVdistributed}  & MADDPG & Centralised & Decentralised & Global & 2D & Synthetic\\
         Wang L. \textit{et al.} \cite{Wang2021_MADDPG}  & MADDPG & Centralised & Decentralised & Global & 2D & Synthetic\\ 
         Liu X. \textit{et al.} \cite{Liu2019UAVqlearning}  & Cluster-based QL & Centralised & Centralised & Global & 3D & Synthetic\\
        Oubbati O. S. \textit{et al.} \cite{Oubbati_2023}  & MADDPG & Centralised & Centralised & Global & 3D & Synthetic\\
         Mozaffari M. \textit{et al.} \cite{Mozaffari2017UAV} & Iterative Search & -- & Centralised & Global & 3D & Synthetic\\
         Omoniwa B. \textit{et al.} \cite{Omoniwa_DQLCI_2021} & Q-Learning & Decentralised & Decentralised & -- & 3D & Synthetic\\ 
         Omoniwa B. \textit{et al.} \cite{Omoniwa2022_Letters} & MAD--DDQN & Decentralised & Decentralised & -- & 3D & Synthetic\\ \hline
         This work & CMAD--DDQN & Decentralised & Decentralised & Local & 3D & Synthetic + \\ 
          &    &   &    &    &   & Real\\ \hline\\
    \end{tabular}
    \vspace{4mm}
    \begin{tabular}{llllll} \hline
        \textit{\textbf{Paper}} & \textit{\textbf{Ground Users}} & 
        \textit{\textbf{CC Partitioning}} & \textit{\textbf{Interference}} & \textit{\textbf{EE}} &  \textit{\textbf{Objective}}\\ \hline \hline
        Liu C. H. \textit{et al.} \cite{Liu2018UAV} & Static & $K$-Cells & \xmark & \cmark & EE, Coverage\\
         Liu C. H. \textit{et al.} \cite{Liu2020UAVdistributed} & Static & $K$-Cells & \xmark & \cmark & EE, Coverage\\ 
         Wang L. \textit{et al.} \cite{Wang2021_MADDPG} & Static & -- & \xmark & \xmark & Energy, Fairness\\
         Liu X. \textit{et al.} \cite{Liu2019UAVqlearning} & Mobile (RW) & $K$-Clusters & \xmark & \xmark & Trajectory\\
         Oubbati O. S. \textit{et al.} \cite{Oubbati_2023} & Mobile (RWP, RPGM) & -- & \cmark & \xmark & Trajectory, Coverage\\
         Mozaffari M. \textit{et al.} \cite{Mozaffari2017UAV} & Static & $K$-Clusters & \cmark & \xmark & Energy\\ 
         Omoniwa B. \textit{et al.} \cite{Omoniwa_DQLCI_2021} & Mobile (RWP) & -- & \cmark & \xmark & Coverage, Energy\\ 
         Omoniwa B. \textit{et al.} \cite{Omoniwa2022_Letters} & Mobile (GMM) & -- & \cmark & \cmark & EE, Outage, Energy\\ \hline  
         This work  & Static + & -- & \cmark & \cmark & EE, Coverage, \\
          & Mobile (RW, RWP, GMM) &  &   &   & Energy, Fairness\\ \hline  
    \end{tabular}
    \label{tab:MARLUAV}\\
\end{table*}

This paper extends the decentralised MARL approach proposed in~\cite{Omoniwa2022_Letters}, where each agent relies on locally sensed information and makes decisions based on implicitly provided neighbour connectivity information reflected in the agent's reward function~\cite{Ming_MARL_1993} (i.e., no communication mechanism was provided on how agents can collaborate to optimise the system's EE in this dynamic network environment). We observed in our previous work~\cite{Omoniwa2022_Letters} that as the number of deployed UAVs in the network increases, a significant drop in the system's EE was observed. The work~\cite{Omoniwa2022_Letters} provided no mechanism for direct collaboration which could significantly impact the agent-controlled UAVs' ability to coordinate while improving the performance of the overall system. The work~\cite{Oubbati_2023} achieved global collaboration among UAVs via a centralised training technique. However, this approach may be impractical with an increase in the number of UAVs deployed. In this work, we aim to achieve collaboration among the agent-controlled UAVs locally via direct communication with the UAVs' neighbours, which in the long run will enhance global coordination while improving the overall system performance. We extend our evaluation to consider real-world data of users' distribution while investigating the impact of various users' mobility models on the overall system's EE. This motivates us to investigate novel collaborative techniques that improve the total system's EE. Hence, we present a collaborative CMAD--DDQN approach, where each agent makes decisions based on its local observation and direct interaction via existing 3GPP guidelines~\cite{3GPPstandard_uav_comm, 3GPPstandard_neighborRelations} with its nearest neighbours.

\begin{table}[!t]
\small
\centering
\caption{Table of Notations}
\label{table:notations}
\begin{tabular}{ll}
  \hline   
 \textit{Notation} & \textit{Definition} \\
  \hline         
        $U$ & Set of quad-rotor UAVs\\
        $\xi$ & Set of ground users\\
        $x_j, y_j, h_j$ & 3D coordinates of UAV $j$\\
        $\gamma$ & Signal-to-interference-plus-noise-ratio\\   
        $\beta$ & Attenuation factor\\
        $\alpha$ & Path loss exponent\\
        $P$ & Transmit power of UAVs\\
        $d$ & distance\\
        $\chi_{int}$ & Set of interfering UAVs\\
        $d_{col}$ & Minimal Collision distance\\
        $\sigma^2$ & Power of the additive white Gaussian noise\\
        $B_w$ & Channel bandwidth\\
        $e_P$ & Propulsion energy\\
        $e_C$ & Communication energy\\
        $e_T$ & Total energy\\     
        $\delta_t$ & Duration of time step\\
        $\kappa_0$ & Coefficient of blade profile power\\
         $\kappa_1$ & Coefficient of induced power\\
        $\kappa_2$ & Coefficient of parasite power\\     
        $U_{tip}$ & Rotor blade's tip speed\\
        $v_0$ & Mean hovering velocity\\
        $\mathcal{R}_j^t$ & Agent $j$'s reward function at time $t$\\
        $C_j^t$ & Agent $j$'s connectivity score at time $t$\\
        $N_d^t$ & Set of distances of neighbouring UAVs\\
        $f_j^t$ & Jain’s fairness index\\
        $e_j^t$ & Agent $j$'s energy level at time $t$\\
        $\mho$ & Agent's cooperative factor\\
        $D_s$ & Dimension of input state space\\
        $D_a$ & Dimension of action space\\
        $s$, $s'$  & State, Next state\\
        $a$ & Action\\
        $r$ & Reward signal\\
        $K$ & Number of hidden layers\\
        $W_k$ & Number nodes in $k^{th}$ layer\\
        $N$ & Number of learning episodes\\
        $T$ & Number of time steps\\
        $\theta, \theta^-$ & Parameters of neural network\\
        $L(\theta)$ & Loss function\\
        $\pi_j$ & Policy of Agent $j$\\
        $\eta$ & Energy efficiency\\
        $E$ &  Bits representing each observation\\
        $U_L$ & Number of UAVs in a neighbourhood\\
        $U_G$ & Number of UAVs globally\\ 
      \hline
 \end{tabular}
 \end{table}

\section{System Model}
\begin{figure}[!t]
\centering
\includegraphics[width=3.5in]{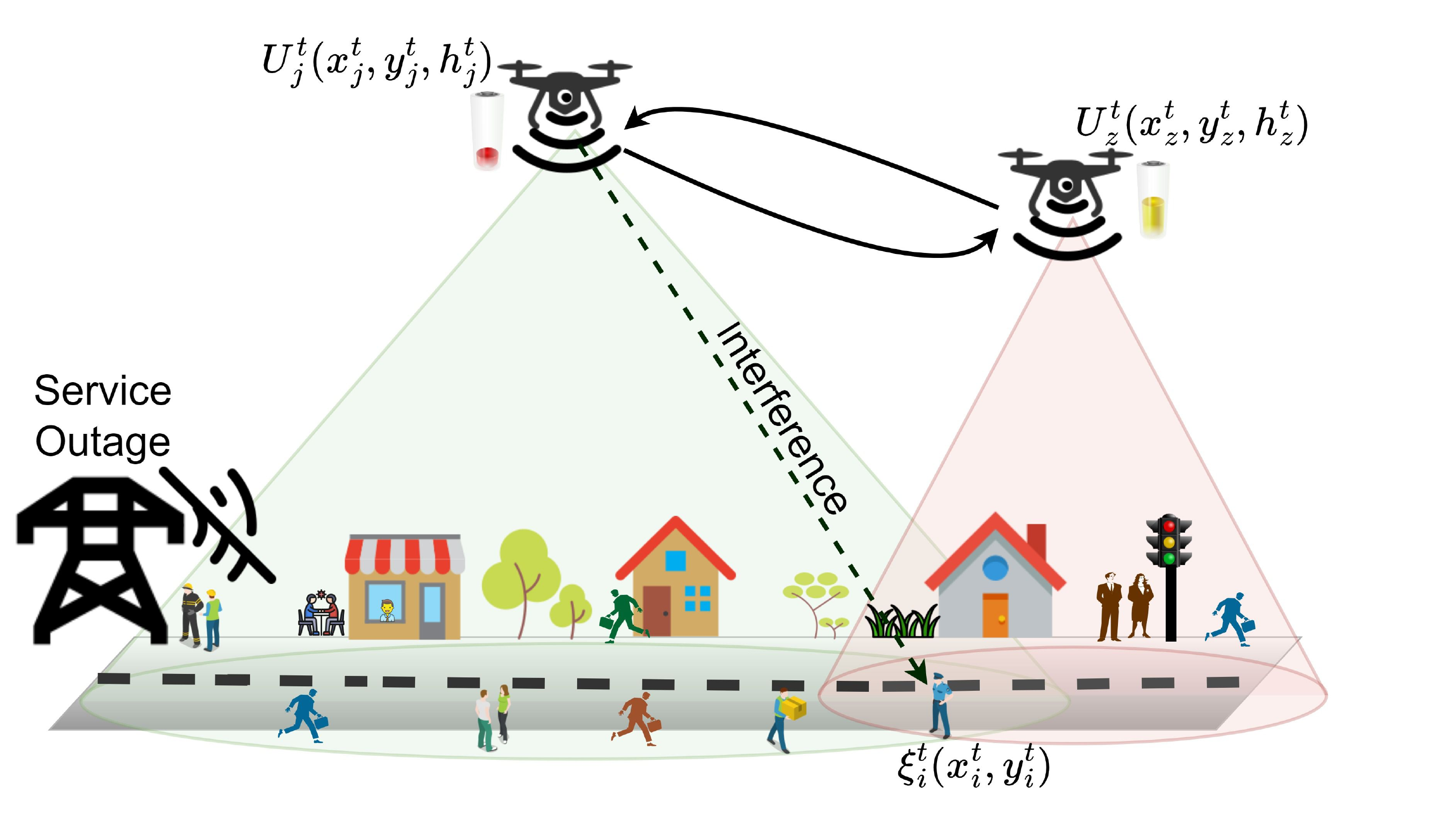}
 \caption{System model for UAVs serving static and mobile ground users. The UAVs directly communicate with nearby neighbours.}
 \label{fig:systemmodel}
\end{figure}

We consider a set of static and mobile ground users $\xi$ located in a given area as in~\cite{Omoniwa2022_Letters}. Each user $i \in \xi$ at time $t$ is located in the coordinate $(x_i^t, y_i^t) \in \mathbb{R}^2$. In this work, We assume connectivity service outages from the existing terrestrial infrastructure due to disasters or increased network load. As such, a set $U$ of quad-rotor UAVs are deployed within the area to provide wireless coverage to the ground users. As an extension to our prior work in~\cite{Omoniwa2022_Letters}, we assume that the UAVs can exchange state information by communicating with nearby neighbours as shown in Figure~\ref{fig:systemmodel}. Table~\ref{table:notations} provides us with the notations used and their definitions.

\subsection{Wireless channel model}
A UAV $j \in U$ providing wireless coverage to ground users at time $t$ is located in the coordinate $(x_j^t, y_j^t, h_j^t) \in \mathbb{R}^3$. Without loss of generality, a guaranteed line-of-sight (LoS) channel condition is assumed, due to the aerial positions of the UAVs in the sky. Each user $i \in \xi$ in time $t$ can be connected to a single UAV $j \in U$ which provides the strongest downlink signal-to-interference-plus-noise-ratio (SINR). SINR is a measure of signal quality. It can be defined as the ratio of the power of a certain signal of interest and the interference power from all the other interfering signals plus the noise power. The SINR between two communicating devices will typically decrease as the distance between the devices increases and it is also a function of the signal propagation and interference in the environment. In particular, the scenario presented in this paper considers the mobility of both UAVs and ground users, which typically causes SINR variation in the propagation environment. An expression for the SINR at time $t$ is given by~\cite{Omoniwa_DQLCI_2021},
\begin{equation}\label{eq:sinr}
    \gamma_{i,j}^t = \frac{\beta P (d_{i,j}^t)^{-\alpha}}{\Sigma_{z \in \chi_{int}} \beta P (d_{i,z}^t)^{-\alpha} + \sigma^2},
\end{equation}
where $\alpha$ and $\beta$ are the path loss exponent and attenuation factor that characterises the wireless channel, respectively. $\sigma^2$ is the power of the additive white Gaussian noise at the receiver, $d_{i,j}^t$ is the distance between the $i$ and $j$ at time $t$, and expressed as $d_{i,j}^t = \sqrt{(x_i^t - x_j^t)^2 + (y_i^t - y_j^t)^2 + (h_j^t)^2}$, while $d_{i,z}^t$ is the distance between the $i$ and $z$ at time $t$, and given as $d_{i,z}^t = \sqrt{(x_i^t - x_z^t)^2 + (y_i^t - y_z^t)^2 + (h_z^t)^2}$ .~$\chi_{int}$ is the set of interfering UAVs within the coverage area (i.e., the operation area of the UAVs). $z$ is the index of an interfering UAV in the set~$\chi_{int}$. $P$ is the transmit power of the UAVs. We assume that the UAVs should keep a minimal distance~$d_{col}$ from each other to avoid collision~\cite{Wang2021_MADDPG}. We model the mobility of mobile users using the random walk (RW), random waypoint (RWP) and the Gauss–Markov (GMM) mobility models~\cite{Song2010, Camp2002}, which allows users to dynamically change their positions. UAVs are expected to optimise their flight trajectory to provide ubiquitous connectivity to users which may be static or mobile. Given a channel bandwidth $B_w$, using Shannon's equation the receiving data rate of a ground user can be expressed~\cite{Galkin2020UAVDeepLearning},
\begin{equation}\label{eq:shannon}
  \mathcal{R}_{i,j}^t = B_w\log_2(1 + \gamma_{i,j}^t).
\end{equation}

\subsection{Connectivity model}
We consider an interference-limited system where coverage is affected by the SINR. Thus, the connectivity score of a UAV $j \in U$ at time $t$ is calculated as~\cite{Liu2020UAVdistributed},
\begin{equation}\label{eq:avgcoveragescore}
C_j^t = \sum_{\forall i \in \xi}w_j^t(i),
\end{equation}
where $w_j^t(i) \in [0, 1]$ represents whether user $i$ is connected to UAV $j$ at time $t$. $w_j^t(i) = 1$ if $\gamma_{i,j}^t > \gamma_{th}$, otherwise $w_j^t(i) = 0$, where $\gamma_{th}$ is the SINR predefined threshold. Likewise $\mathcal{R}_{i,j}^t = 0$ if user $i$ is not connected to UAV $j$. In a variety of network service scenarios, including disasters, it is desirable to have nearly all ground users connected fairly to the available UAVs. As such, we define geographical fairness using the Jain’s fairness index as~\cite{Liu2020UAVdistributed, Omoniwa_DQLCI_2021, Wang2021_MADDPG},
\begin{equation}\label{eq:JainFairness}
f_j^t = \frac{\big(\sum_{\forall j \in N}C_t(j)\big)^2}{N \sum_{\forall j \in N}C_t(j)^2}
\end{equation}

\subsection{Energy consumption model}
During a flight operation, UAV $j \in U$ at time $t$ expends energy $e_j^t$. A UAV's total energy $e_T$ is expressed as the sum in propulsion~$e_P$ and communication~$e_C$ energies, $e_T = e_P + e_C$. Since $e_C$ is practically much smaller than $e_P$, i.e., $e_C \ll e_P$ \cite{Omoniwa_DQLCI_2021}, we ignore $e_C$. Therefore, we consider the propulsion power consumption model for a rotary-wing UAV used in \cite{Omoniwa2022_Letters}. A closed-form analytical propulsion power consumption model for a rotary-wing UAV at time $t$ is given as \cite{Zeng2019UAV},
\vspace{-2mm}
\begin{equation}\label{eq:powertime}
P(t) = \kappa_0 \Big(1 + \frac{3V^2}{U^2_{tip}}\Big)  + \kappa_1 \Big( \sqrt{1 + \frac{V^4}{4v_0^4}} + \frac{V^2}{2v_0^2}\Big)^{\frac{1}{2}} + \frac{\kappa_2}{2} V^3,
\end{equation}
where $\kappa_0$, $\kappa_1$ and $\kappa_2$ are the flight constants of the UAVs (e.g., disk area, rotor radius, drag ratio, weight, air density), $U_{tip}$ is the tip speed of the rotor blade, $v_0$ is the mean hovering speed, and $V$ is the UAVs' speed at time $t$. In particular, we take into account the basic operations of the UAV, for instance, hovering and accelerating. Therefore, we can derive the average propulsion power over all time steps as $\frac{1}{T}\sum_{t=1}^{T}P(t)$,  and the total energy consumed by UAV $j$ at time $t$ is given as~\cite{Omoniwa_DQLCI_2021},
\vspace{-1mm}
\begin{equation}\label{eq:propenergy}
%e_{j}^t = \delta_t \sum_{t=1}^{t}P(t),
e_{j}^t = \delta_t \cdot P(t),
\vspace{-1mm}
\end{equation}
where $\delta_t$ is the duration of each time-step. The EE of UAV $j$ can be expressed as the ratio of the data throughput and the energy consumed in time-step $t$, expressed as~\cite{Omoniwa2022_Letters},
\begin{equation}\label{eq:energyEfficiency}
    \eta_j^{t} = \frac{\sum\limits_{i \in \xi} \mathcal{R}_{i,j}^t}{e_j^t}.
\end{equation}
Maximisation of the number of connected users $C_j^t$ will maximise the total amount of data $\sum\limits_{i \in \xi} \mathcal{R}_{i,j}^t$ the UAV $j$ will deliver in time-step $t$ which, for a given amount of consumed energy $e_j^t$, will also maximise the total EE $\eta_{tot}$. Therefore, the total systems' EE over all time-step is expressed as,
\begin{equation}\label{eq:tot_energyEfficiency}
    %%\eta_{tot} =~ \sum\limits_{t=1}^T \sum\limits_{j \in U} \sum\limits_{i \in v} \mathbb{R}_{i,j}^t \Bigg/ \Bigg(\sum\limits_{t=1}^T \sum\limits_{j \in U} e_j^t \Bigg).   
    \eta_{tot} =~ \frac{\sum\limits_{t=1}^T \sum\limits_{j \in U} \sum\limits_{i \in \xi} \mathcal{R}_{i,j}^t}{\sum\limits_{t=1}^T \sum\limits_{j \in U} e_j^t}.
\end{equation}
%%%%%%%%%%%%%%%%%%%%%%%%%%%%%%%%%%%%%
\section{Multi-Agent Reinforcement Learning Approach for Energy Efficiency Optimisation}
\label{section:mad-dqn}
In this section, the problem is formulated. We go further to present our proposed CMAD--DDQN algorithm to improve the trajectory of each UAV in a manner that maximises the total system's EE.

\vspace{-2mm} 
\subsection{Problem Formulation} 
Our objective is to maximise the total system's EE by jointly optimising its 3D trajectory, the number of connected users, and the UAVs' energy consumed while deployed to serve ground users under a strict energy budget. Therefore, we formulate the optimisation problem as~\cite{Omoniwa2022_Letters}, 
\begin{subequations}
\begin{align} \label{eq:problem_statement}
&\underset{\forall j \in N:~\mathbf{x_j^t,~y_j^t,~h_j^t,~e_j^t,~C_j^t}}{\max}~\eta_{tot}\\
\label{eq:problem_statement_b}
\text{s.t.} \quad &  \gamma_{i,j}^t \geq \gamma_{th},\quad \forall w_j^t(i) \in [0, 1],~i,~j,~t, \\ \label{eq:problem_statement_c}
&e_j^t \leq e_{\max},\quad \quad \quad \quad \quad \quad \quad \forall j,~t,\\ \label{eq:problem_statement_d}
&x_{\min}~\leq~x_j^t~\leq x_{\max},\quad \quad ~~\forall j,~t,\\ \label{eq:problem_statement_e}
&y_{\min}~\leq~y_j^t~\leq y_{\max},\quad \quad ~~~\forall j,~t,\\
\label{eq:problem_statement_f}
&h_{\min}~\leq~h_j^t~\leq h_{\max},\quad \quad ~~~\forall j,~t,
\end{align}
\end{subequations}
where $x_{min}$, $y_{min}$, $h_{min}$ and $x_{max}$, $y_{max}$, $h_{max}$ are the minimum and maximum coordinates of $x$, $y$ and $h$, respectively. $e_{\max}$ is the UAV's maximum energy budget. The constraints in Equation (\ref{eq:problem_statement_b})--(\ref{eq:problem_statement_f}) ensure that the UAVs stay within tolerable bounds. The constraint~(\ref{eq:problem_statement_b}) is to ensure that all users always meet the minimum SINR threshold. The constraint in Equation~(\ref{eq:problem_statement_c}) ensures that the UAVs do not exceed their maximum energy budget, while the constraints in Equation (\ref{eq:problem_statement_d})~--~(\ref{eq:problem_statement_f}) are to keep the UAVs within the operating area. When multiple UAV small cells or access points share the same frequency spectrum and are deployed in a shared environment, managing interference in the network becomes increasingly difficult. The amount of interference received from other interfering sources is a function of the UAVs' locations~\cite{Mozaffari2017UAV}. The complexity of the problem (\ref{eq:problem_statement}) increases as more UAVs are deployed in a shared wireless environment, and belong to a well-known class of NP-hard problems~\cite{Liu2019UAVqlearning}. In particular, multi-agent problems of this nature have been proven to be very difficult as more than 2 agent-controlled UAVs are deployed in a shared environment~\cite{Bernstein_2002}. Hence, it is difficult to solve using conventional optimization approaches~\cite{Omoniwa2022_Letters, Liu2019UAVqlearning}. Under dynamic network conditions, UAVs may often prioritise their individual goals over the collective goal of improving the total system's EE. This selfish behaviour is due to non-stationary induced by the agents' policies. In such cases, cooperative MARL approaches may be suitable when there is a conflict between the individual and collective interests of UAVs. Deep RL has been proven to perform well in decision-making tasks in this kind of dynamic environment~\cite{Zhang2021uavBS}. Intuitively, a reduction in the complexity can be achieved when the agent-controlled UAVs fully share their observations with their neighbours in each time step~\cite{Szer2004_Communication}. However, sharing all observations may result in increased communication overhead. Nevertheless, the performance gain that may be derived via communication may outweigh this drawback. In our prior work~\cite{Omoniwa2022_Letters} where UAVs have no mechanism to communicate, the system's EE degraded as the number of UAVs in the network increased. Hence, we adopt a cooperative communication-enabled deep MARL approach to solve the system's EE optimisation problem.

%%\begin{comment}
\begin{algorithm}
\footnotesize
\caption{Double Deep Q-Network (DDQN) for Agent $j$ with Direct Collaboration with its Neighbours}\label{Algorithm_DDQN}
\begin{algorithmic}[1]
%\State \textbf{Initialize:}
\State Input: UAV3Dposition $(x_j^t,~y_j^t,~h_j^t)$, ConnectivityScore $c^t_j$, InstantaneousEnergyConsumed $e^t_j$, UAVneighbourDistances $N_d^t$, NeighboursConnectionScore $c^t_z$, NeighboursInstantaneousEnergyConsumed $e^t_z$ $\in S$ 
\State Output: Q-values corresponding to each possible action~$(+x_s,~0,~0)$, $(-x_s,~0,~0)$, $(0,+y_s,~0)$, $(0,-y_s,~0)$, $(0,~0,+z_s)$, $(0,~0,-z_s)$, $(0,~0,~0)$~$\in A_j$.
\ForAll{$a \in A_j$ and~$s \in S$}:
\State \parbox[t]{0.8\linewidth}{$Q_{(1)}(s, a)$, $Q_{(2)}(s, a)$, $\mathcal{D}$ -- empty replay buffer,  $\theta$ -- initial network parameters, $\theta^{-}$ -- copy of $\theta$, $N_r$ -- maximum size of replay buffer, $N_b$ -- batch size, $N^{-}$ -- target replacement frequency.}
\State $s$ $\gets$ initial state
\State 1500 $\gets$ maxStep
\While{\emph{goal} not Reached \& Agent \emph{alive} \& maxStep not reached} 
\State \emph{s} $\leftarrow$ MapLocalObservationToState(\emph{Env}) 
\State \Comment{\parbox[t]{0.8\linewidth}{\texttt{Execute $\epsilon$-greedy method based on $\pi_j$}}}
\State \emph{a} $\leftarrow$ DeepQnetwork.SelectAction(\emph{s})
\State \Comment{\parbox[t]{0.8\linewidth}{\texttt{Agent executes action in state $s$}}}
\State  \emph{a}.execute(\emph{Env})
\If { \emph{a}.execute(\emph{Env}) is True}
\State \Comment{\parbox[t]{0.8\linewidth}{\texttt{Map sensed observations to new state $s'$}}}
\State Env.UAV3Dposition
\State \Comment{\parbox[t]{0.8\linewidth}{\texttt{using Equation (\ref{eq:avgcoveragescore})}}}
\State Env.ConnectivityScore
\State \Comment{\parbox[t]{0.8\linewidth}{\texttt{using Equation (\ref{eq:propenergy})}}}
\State Env.InstantaneousEnergyConsumed 
\State \Comment{\parbox[t]{0.8\linewidth}{\texttt{Map observations exchanged from closest neighbours based on an existing ANR mechanism for UAV communication to new state $s'$}}}
\State Env.Neighbour.UAVneighbourDistances
\State Env.Neighbour.ConnectionScore 
\State Env.Neighbour.InstantaneousEnergyConsumed 
\EndIf
\State \Comment{\parbox[t]{0.8\linewidth}{\texttt{using Equation (\ref{eqnreward})}}}
\State \emph{r} $\leftarrow$ Env.RewardWithCooperativeNeighbourFactor 
\State \Comment{\parbox[t]{0.8\linewidth}{\texttt{ Execute UpdateDDQNprocedure()}}}
\State Sample a minibatch of $N_b$ tuples $(s, a, r, s') \sim Unif(\mathcal{D})$ \label{Algline:dqn_start}
\State Construct target values, one for each of the $N_b$ tuples:
\State Define $a^{max} (s'; \theta) =  \arg \max_{a'} Q_{(1)}(s', a'; \theta)$
\If {\emph{$s'$} is Terminal}
\State $y_j =  r$
\Else
\State $y_j =  r + \gamma Q_{(2)}(s', a^{max} ((s'; \theta); \theta^{-})$
\EndIf
\State Apply a gradient descent step with loss $\parallel y_j - Q(s, a; \theta)\parallel^2$
\State Replace target parameters $\theta^{-} \gets \theta$ every $N^{-}$ step \label{Algline:dqn_stop}
\EndWhile
\State \textbf{endwhile}
\EndFor
\end{algorithmic}
\end{algorithm}

%\begin{comment}
\begin{figure*}[!t]
\centering
\includegraphics[width=6.5in]{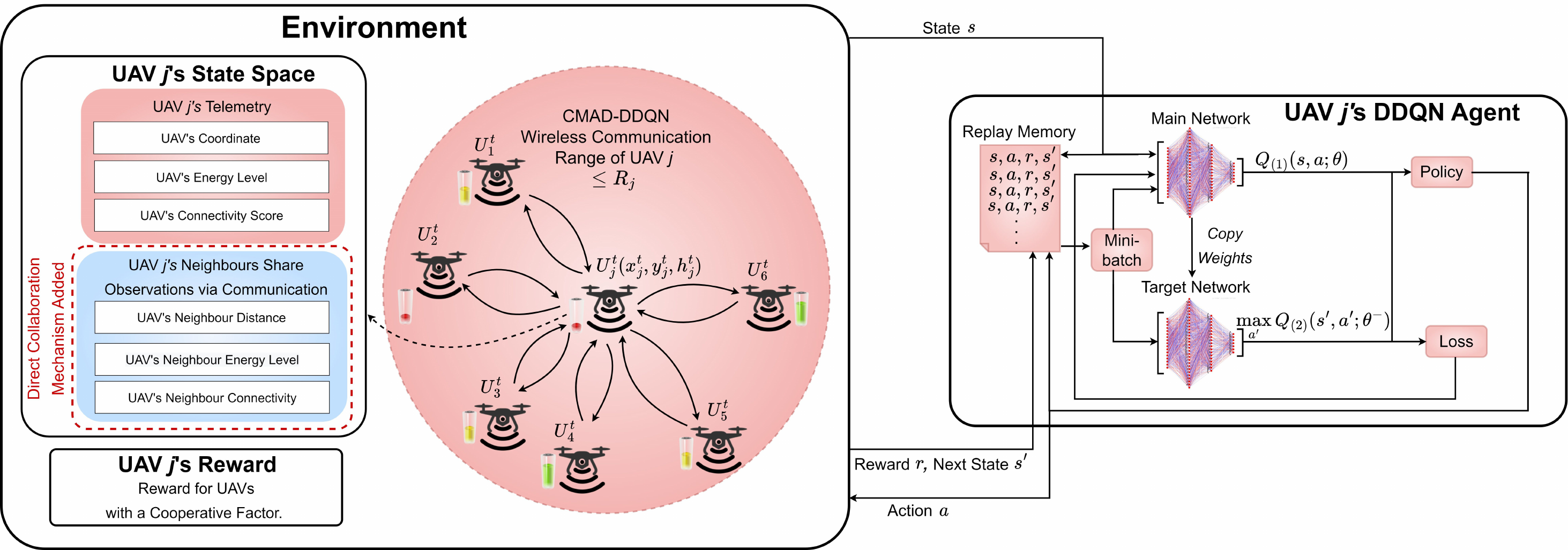}
\caption{Communication-enabled multi-agent decentralised double deep Q-network (CMAD--DDQN) framework where each UAV $j$ equipped with a DDQN agent interacts and shares knowledge with its nearest neighbours which makes up the state space. Each UAV directly collaborates to improve the overall system performance. At each time step, UAV $j$'s DDQN agent observes its present state $s$ in the environment and updates its trajectory by selecting an action $a$ in accordance with its policy, hereby receiving a reward $r$ and transiting to a new state $s'$.}
\label{fig:madddqn}
\end{figure*}
%\end{comment}
 
%%%%%%%%%%%%
\subsection{Cooperative Communication-Enabled Multi-Agent Decentralised Double Deep Q-Network (CMAD--DDQN)}
It is expected that UAVs will be legally required to broadcast their telemetry information for safety reasons, which involves sharing their coordinates, UAV identification, flight plans (or rather velocity and direction, for security and privacy reasons), vehicle type~\cite{Vinogradov2020}. This communication can be done through standardized 3GPP sidelink communication (enabling device-to-device (D2D) communications without going through the network infrastructure). Hence, we propose a cooperative CMAD--DDQN approach that relies on a communication mechanism among neighbouring UAVs for improved system performance. In the scenario we consider, each agent's reward reflects the coverage performance in its neighbourhood. As seen in Figure \ref{fig:madddqn}, each UAV is controlled by a double deep Q-network (DDQN) agent that aims to maximise the system's EE by jointly optimising its 3D trajectory, number of connected ground users, and the energy consumed by the UAVs. We assume that as the agents interact with each other in a shared and dynamic environment, they may observe learning instabilities due to conflicting policies from other agents. Algorithm \ref{Algorithm_DDQN} shows the DDQN for Agent $j$ with direct collaboration with its neighbours. In our proposed approach, the control overhead incurred by each sensory-exchanging agent-controlled UAV per step is bounded by $(U_L - 1) \times E$ (Refer to Case-study 2, \cite{Ming_MARL_1993}), where $U_L$ is the number of UAVs within the neighbourhood,  $E$ is the number of bits required to represent each observation by the agent. On the other hand, the control overhead of a closely related evaluation baseline~\cite{Liu2020UAVdistributed} is bounded by $(U_G - 1) \times E$, where $U_G$ is the overall number of UAVs in the system. Agent-controlled UAVs may require certain information from their nearest neighbours for decision-making. Our closely related evaluation baseline~\cite{Liu2020UAVdistributed} considered energy consumption, UAV positions, flying direction, coverage scores and distance of all UAVs. However, this implies additional overhead since each UAV has a global view of the entire state of other UAVs in the network. Our proposed approach is expected to reduce this overhead while providing UAVs with additional knowledge to improve the overall network performance. As described in Figure \ref{fig:madddqn}, UAV $j$ receives the neighbours' distances, energy levels, and connectivity score from at most 6 closest neighbours within its communication range, which may help to improve its performance locally. Agent $j$ executes an action $a$ in its present state $s$ by following an $\epsilon$--greedy policy. The agent then transits to a new state $s'$ and receives a reward that reflects the coverage performance in its neighbourhood as given in (\ref{eqnreward}). Furthermore, the DDQN procedure described on line \ref{Algline:dqn_start}--\ref{Algline:dqn_stop} optimises the agent's decisions. We explicitly define the states, actions, and reward as follows.
\subsubsection{State space}
We consider the three-dimensional (3D) position of each UAV~\cite{Liu2019UAVqlearning}, the UAV's connectivity score, the UAV's instantaneous energy level, maximum of 6 closest neighbour distances using a defined communication mechanism, the neighbour connectivity score, and neighbour instantaneous energy consumed at time $t$, expressed as a tuple, \textlangle{}$x^t : \{0, ..., x_{max}\},~y^t : \{0, ..., y_{max}\},~h^t : \{h_{min}, ..., h_{max}\},~C_j^t,\\~e_j^t,~N_d^t, ~C_z^t, ~e_z^t$\textrangle{}. Our rationale for considering a maximum of 6 neighbouring UAVs stems from the Honeycomb configuration of multiple base stations, which is widely applied in mobile communications.
\subsubsection{Action space}
At each time-step~$t \in T$, each UAV executes an action by changing its direction along the 3D coordinates. Unlike a closely related work and evaluation baseline~\cite{Liu2020UAVdistributed}, we discretise the agent's actions following the design from ~\cite{Omoniwa2022_Letters, Liu2019UAVqlearning}, as follows:~$(+x_s,~0,~0)$, $(-x_s,~0,~0)$, $(0,+y_s,~0)$, $(0,-y_s,~0)$, $(0,~0,+z_s)$, $(0,~0,-z_s)$ and $(0,~0,~0)$. Our rationale to discretise the action space was to ensure that the agents quickly adapt and converge to an optimal policy.
\subsubsection{Reward}
The objective of each agent-controlled UAV is to learn a policy that maximises the system's EE by jointly maximising the connectivity of ground users while minimising the total UAV's energy consumption. We formulate the reward for each agent $j$ in each time-step $t \in T$ by introducing a shared cooperative factor, denoted by~$\mho$, to drive the agent-controlled UAVs towards cooperative behaviours. This is given by~\cite{Omoniwa2022_Letters},
    \begin{equation}\label{eqnreward}
        \mathcal{R}_j^t =
        \begin{cases}
          ~~\mho + \omega + 1, & \text{if}\ C_j^t > ~C_j^{t - 1}\\%and~$C_t(k)$\\
          ~~\mho + \omega, & \text{if}\ C_j^{t} = ~C_j^{t - 1}\\
          ~~\mho + \omega - 1, & \text{otherwise,}
        \end{cases}
        \vspace{-1mm}
    \end{equation}
where $C_j^{t}$ and $C_j^{t - 1}$ are the connectivity score in the present and previous time-step, respectively. $\omega~=~\frac{e_j^{t - 1} - e_j^{t}}{e_j^{t} + ~e_j^{t - 1}}$, where $e_j^{t}$ and $e_j^{t - 1}$ are the instantaneous energy consumed by agent $j$ in present and previous time-step, respectively. To enhance cooperation, we assign each agent a `$+1$' incentive from its neighbourhood via a function~$\mho$ only when the overall connectivity score, which is the total number of connected users by UAVs in its neighbourhood in the present time-step $C^{t}_o$ exceeds that in the previous time-step $~C^{t - 1}_o$, otherwise, the agent receives a `$-1$' incentive. We compute~$\mho$ as~\cite{Omoniwa2022_Letters},
    \begin{equation}\label{coop_factor}
    \mho =
    \begin{cases}
      ~+1, & \text{if}\ C^{t}_o~> ~C^{t - 1}_o\\%and~$C_t(k)$\\
      ~-1, & \text{otherwise.}
    \end{cases}
    \end{equation}

\subsection{DDQN Implementation}
The neural network (NN) architecture of Agent $j$'s DDQN shown in Figure \ref{fig:madddqn} comprises a 23-dimensional input vector state space, 2 densely connected layers with 128 and 64 nodes, with each using a rectified linear unit (ReLU) activation function, and finally yielding an output layer with 7 dimensions. Our decentralised approach assumes agents to be independent learners while relying on direct collaboration with nearby UAVs.

\begin{theorem} \label{theorem:ddqn}
The time complexity of the decentralised double deep Q-network algorithm is approximate $\mathcal{O}\Big(NT (D_sW_1 + \sum_{k=1}^{K} W_k W_{k+1})\Big)$.
\end{theorem}
\begin{pot}%[Proof of Theorem \ref{theorem:ddqn}]
Algorithm \ref{Algorithm_DDQN} shows that after $T$ time steps and $N$ learning episodes, the neural network parameters of the DDQN algorithm converge and tend to be stable. The time complexity of a neural network (NN) represents the number of operations of the network model, which is determined by the dimension of input state $D_s$ and action $D_a$, the number of layers and the number of neurons in each layer of the NN~\cite{Tan2022_time_complexity}. The operation times of the DDQN in each time step can be expressed as $\mathcal{O}\Big(D_sW_1 + \sum_{k=1}^{K} W_k W_{k+1}\Big)$, where $K$ is the number of hidden layers of the NN, and $W$ is the number nodes in each layer. Hence, the time complexity of the decentralised double deep Q-network algorithm is approximate $\mathcal{O}\Big(NT \big(D_sW_1 + \sum_{k=1}^{K} W_k W_{k+1}\big)\Big)$. The time complexity of a closely related work and evaluation baseline~\cite{Liu2020UAVdistributed} (MADDPG) is approximately $\mathcal{O}\Big(NT \big(D_sW_1 + \sum_{k=1}^{K} W_k W_{k+1}\big)\Big) + \mathcal{O}\Big(NT\big((D_a + D_s)W_1 + \sum_{k=1}^{K} W_k W_{k+1}\big)\Big)$.
\end{pot}

%%%%%%%%
During the training phase and given the state information as input, Agent $j$ trains the main network to improve its decisions by yielding Q-values that match each possible action as output. The maximum Q-value obtained is a determinant of the action the agent executes. At each time-step Agent $j$ observes its present state $s$ and updates its trajectory by selecting an action $a$ according to its policy. Following its action in time-step $t$, Agent $j$ observes a reward $r$ which is defined in (\ref{eqnreward}), and transits to a new state $s'$~\cite{Sutton1998}. The information $(s,a,r,s')$ is inputted in the replay memory as shown in Figure \ref{fig:madddqn}. Agent $j$ now samples the random mini-batch from the replay memory and uses the mini-batch to get $y_j$. The optimisation is performed with $L(\theta)$ and $\theta$ updated accordingly. The target Q-network updates the parameters $\theta^-$ with the same parameters $\theta$ of the main network in every 100th time-step. We set the memory size was set to 10,000 for the training while using a mini-batch size of 1024. We perform the optimisation using a variant of the stochastic gradient descent called RMSprop to minimise the loss following the methodology described in \cite[Chapter 4]{Francois-Lavet2018}. The learning rate is set to 0.0001 and the discount factor of 0.95 was applied. Our Q-networks were trained by running multiple episodes, and at each training step the $\epsilon$-greedy policy is used to have a balance between exploration and exploitation~\cite{Francois-Lavet2018}. In the $\epsilon$-greedy policy, the action is randomly selected with $\epsilon$ probability, whereas the action with the largest action value is selected with a probability of $1-\epsilon$~\cite{Omoniwa2022_Letters}. The initial value of $\epsilon$ was set to 1 and linearly decreased to 0.01.

\begin{table}[!t]
\small
\centering
\caption{Simulation Parameters}
\label{table:parameters}
\begin{tabular}{ll}
  \hline
 \textit{Parameters} & \textit{Value} \\
  \hline
  Software platform & Python\\
        Library & Pytorch\\
       Optimiser & RMSprop\\
       Loss function &  MSELoss\\
       Learning rate~\cite{Omoniwa2022_Letters, Qu_UAV_2022} & 0.0001\\
       Discount factor~\cite{Omoniwa2022_Letters} & 0.95\\
       Hidden layers~\cite{Omoniwa2022_Letters} & 2 (128, 64)\\ 
       Activation function~\cite{Omoniwa2022_Letters} & ReLu\\ 
       Policy~\cite{Omoniwa2022_Letters, Liu2019UAVqlearning} & $\epsilon$-greedy\\  
       Memory size~\cite{Omoniwa2022_Letters} & 10,000\\ 
       Batch size~\cite{Omoniwa2022_Letters} & 1024\\ 
       Episodes & 250\\
       Max. steps per episode & 1500\\
       No. of ground users & 400\\
       Mobility Models & GMM, RW, RWP\\
       Ground user direction & [0, 2$\pi$]\\ 
       Pedestrian Velocity~\cite{Omoniwa2022_Letters} & [0, 2]~m/s\\ 
       UAV Velocity~\cite{Omoniwa2022_Letters} & [0, 20]~m/s\\ 
       Number of UAVs & [2--12]\\
       Weight per UAV~\cite{Omoniwa2022_Letters} & 16 kg\\
       $\kappa_0$~\cite{Omoniwa2022_Letters} & 79.85~J/s\\
        $\kappa_1$~\cite{Omoniwa2022_Letters} & 88.63~J/s\\
        $\kappa_2$~\cite{Omoniwa2022_Letters} & 0.018~kg/m\\  
        $U_{tip}$~\cite{Omoniwa2022_Letters}& 120~m/s\\
        $v_0$~\cite{Omoniwa2022_Letters} & 4.03~m/s\\
       Nominal battery capacity~\cite{Omoniwa2022_Letters} & 16,000 mAh\\ 
       Maximum transmit power~\cite{Liu2019UAVqlearning} & 20 dBm\\ 
       Noise power~\cite{Mozaffari2017UAV} & -130 dBm\\
       SINR threshold~\cite{Omoniwa2022_Letters, Mozaffari2017UAV} & 5 dB\\
       Bandwidth~\cite{Liu2019UAVqlearning} & 1 MHz\\
       Pathloss exponent~\cite{Mozaffari2017UAV, Liu2019UAVqlearning} & 2\\
       UAV step distance ($\forall~x_s, y_s, z_s$)~\cite{Omoniwa2022_Letters} & [0--20] m\\
       UAV Collision distance $d_{col}$~\cite{Omoniwa2022_Letters} & 20 m\\
      \hline
 \end{tabular}
 \end{table}

\begin{figure}[t]
\centering
\includegraphics[width=3.3in]{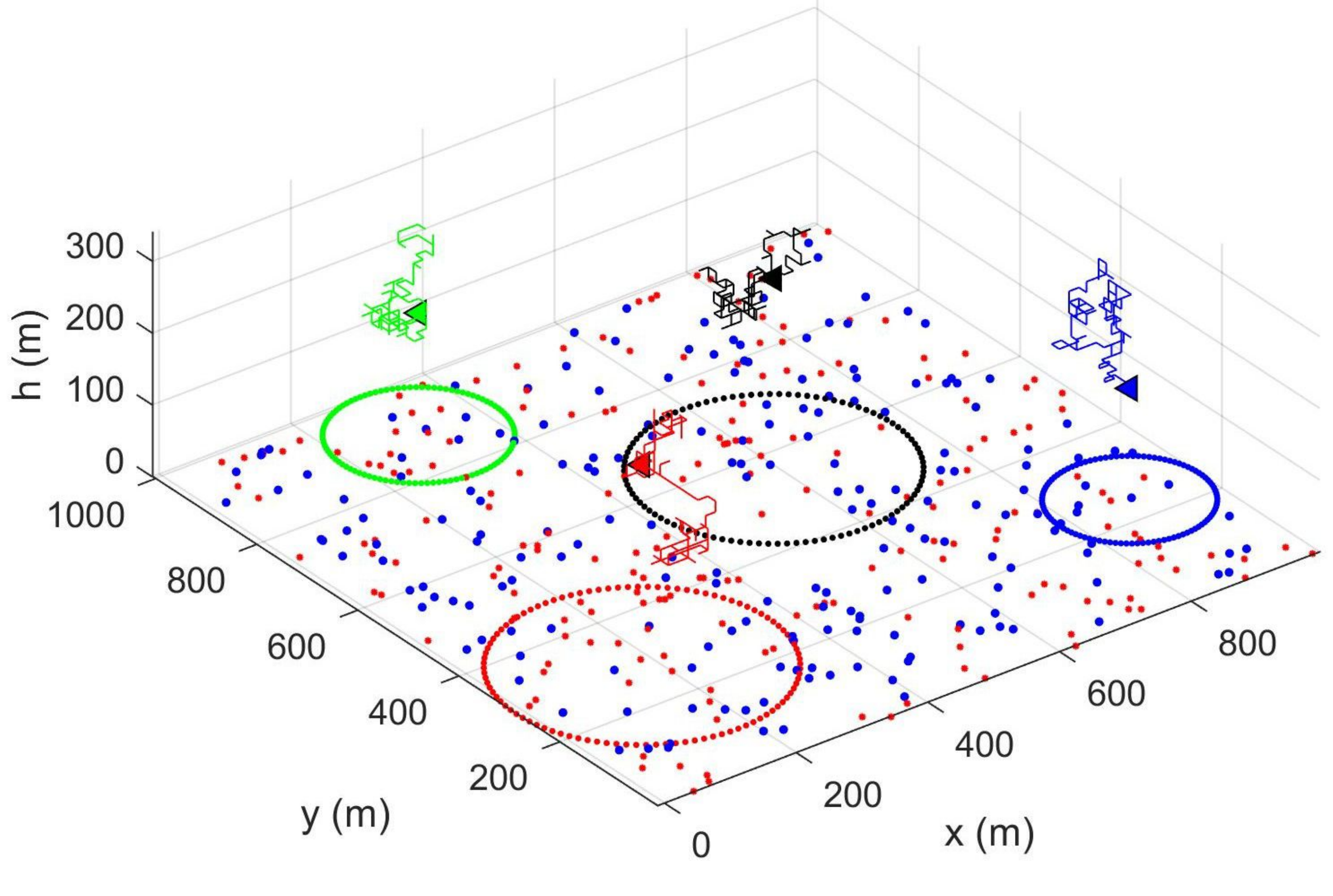}
 \caption{Simulation snapshot with flight trajectories of 4 UAV base stations serving static (blue) and mobile (red) ground users in a 1000$\times$1000 $m^2$ area.}
 \label{fig:snap}
\end{figure}

\section{Evaluation} \label{Evaluation_Plan}
\subsection{Simulation Settings}
The problem formulation described in Section \ref{section:mad-dqn} is simulated using the Python library, \textit{Pytorch}. Simulation parameters are presented in Table~\ref{table:parameters}. We deploy a varying number of UAVs ranging from 2 to 12 to serve both static and mobile ground users in a 1000$\times$1000 $m^2$ area as shown in the environment on Figure \ref{fig:snap}. We perform several runs of experiments and gathered results from independent runs of trained episodes. 

\subsection{Baselines}
In this work, we compare the effectiveness of the proposed CMAD--DDQN approach against the following baselines: \begin{itemize}[leftmargin=*]
    \item The random policy, where UAVs choose their flight directions and travel distances randomly at each time-step~$t$.
    \item The MADDPG~\cite{Liu2020UAVdistributed} approach that considers a 2D trajectory optimisation while neglecting interference from nearby UAV cells. Here, the action space for each agent is continuous.
    \item The MAD--DDQN~\cite{Omoniwa2022_Letters} approach considers a 3D trajectory optimisation and the interference from nearby UAV cells while neglecting communication among UAVs in a shared network environment.
\end{itemize}

\subsection{Metrics}
We considered the following metrics for performance evaluation:
\begin{itemize}[leftmargin=*]
    \item The total energy efficiency ($\eta_{tot}$)~\cite{Omoniwa2022_Letters, Galkin2022fixedWingUAV}: This is defined as the ratio of the total throughput and the total energy consumed given as Equation~(\ref{eq:energyEfficiency}). This metric gives us an insight into how much energy is expended by the UAVs to deliver certain bits of information. It is desirable to improve the total system's EE since we want as much information as possible to be delivered using a minimum amount of energy.
    \item The number of connected ground users~\cite{Omoniwa2022_Letters, Omoniwa_DQLCI_2021}: This metric shows the total number of ground users connected to UAV small cells. We express this as a percentage. It gives an estimate of how well the coverage performance of the UAVs is. A low value in the number of connected users indicates poor coverage by the UAVs, hence it is desirable to have a high number of connected users.
    \item The total energy consumed by UAVs~\cite{Omoniwa2022_Letters, Omoniwa_DQLCI_2021}: This metric is very important as it directly impacts battery life and a UAV that completely depletes its battery dies out and can not provide coverage any longer.
    \item The fairness index~\cite{Liu2020UAVdistributed, Omoniwa_DQLCI_2021, Wang2021_MADDPG}: This metric reflects the QoS level of ground users served by UAVs from the initial time-step to the current time-step given as Equation (\ref{eq:JainFairness}). All ground users need to be fairly served, being connected to a UAV small cell as much as possible.
\end{itemize}  

\begin{figure*}
     \centering
     \begin{subfigure}[b]{0.3\textwidth}
         \centering
         \includegraphics[width=\textwidth]{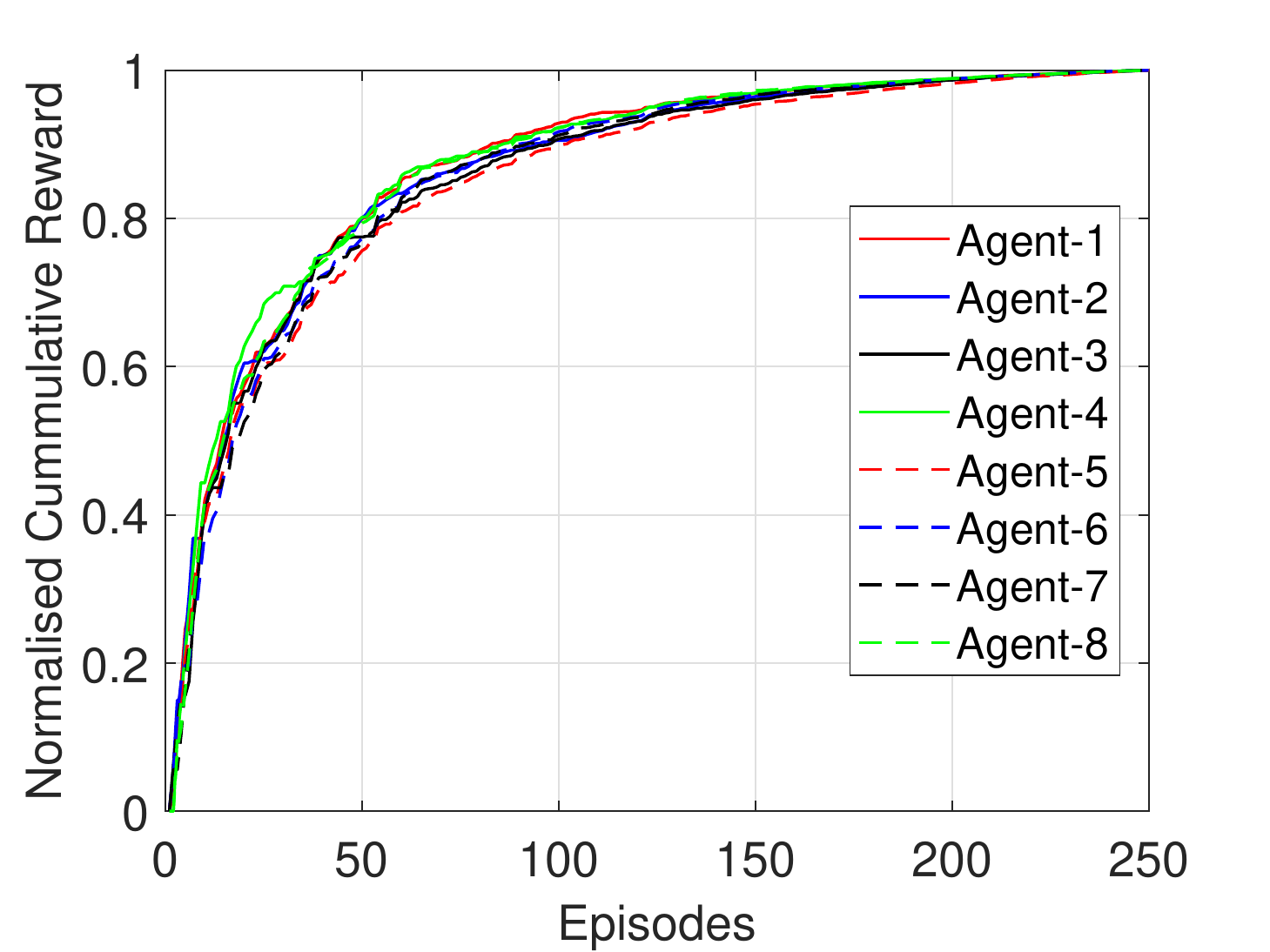}
         \caption{Cumulative reward per agent-controlled UAV vs. episodes.}
         \label{subfig:RewVsepi_cmad}
     \end{subfigure}
     \hfill
     \begin{subfigure}[b]{0.3\textwidth}
         \centering
         \includegraphics[width=\textwidth]{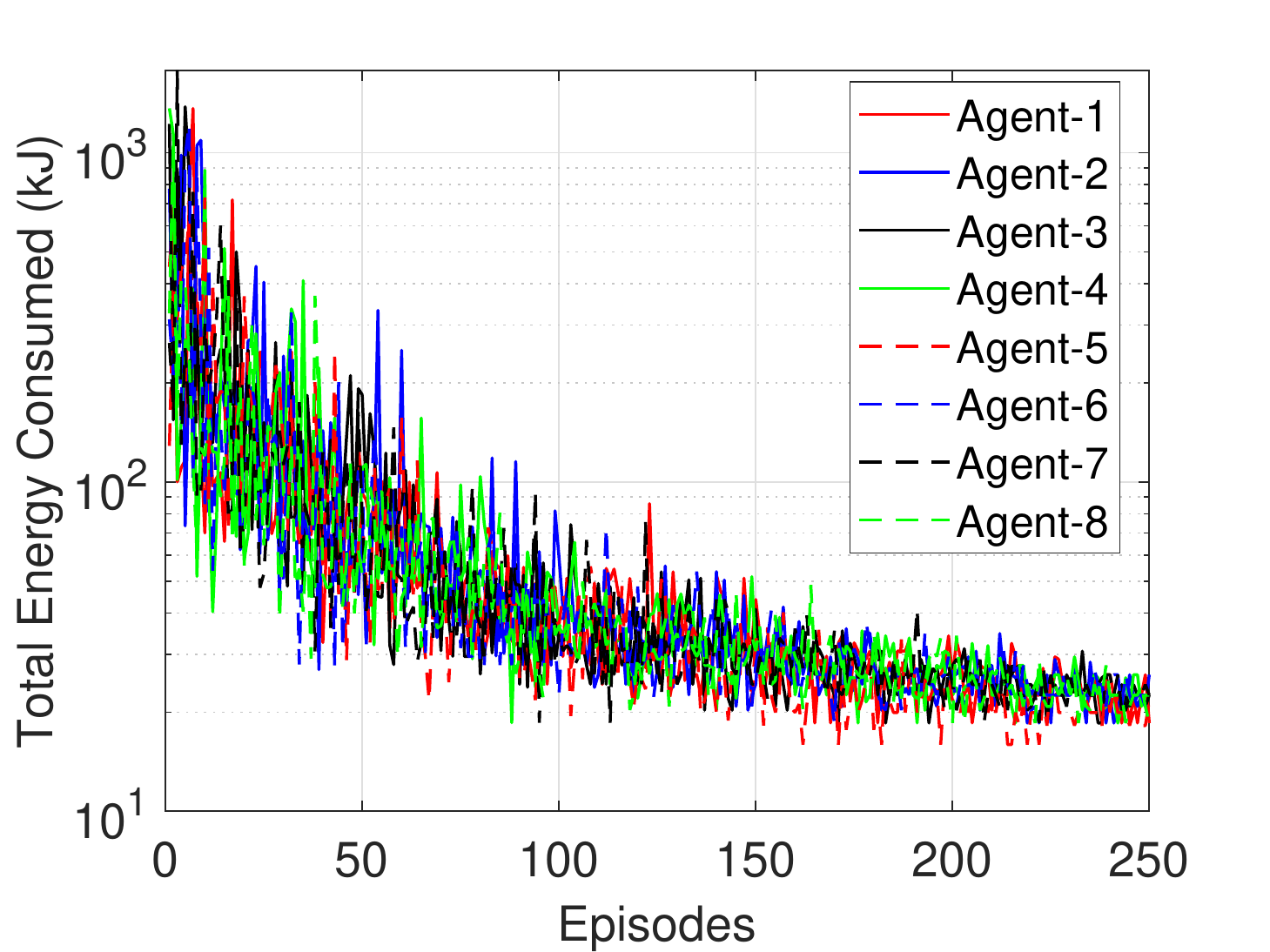}
         \caption{Total energy consumed per agent-controlled UAV vs. episodes.}
         \label{subfig:eneVsepi_cmad}
     \end{subfigure}
     \hfill
     \begin{subfigure}[b]{0.3\textwidth}
         \centering
         \includegraphics[width=\textwidth]{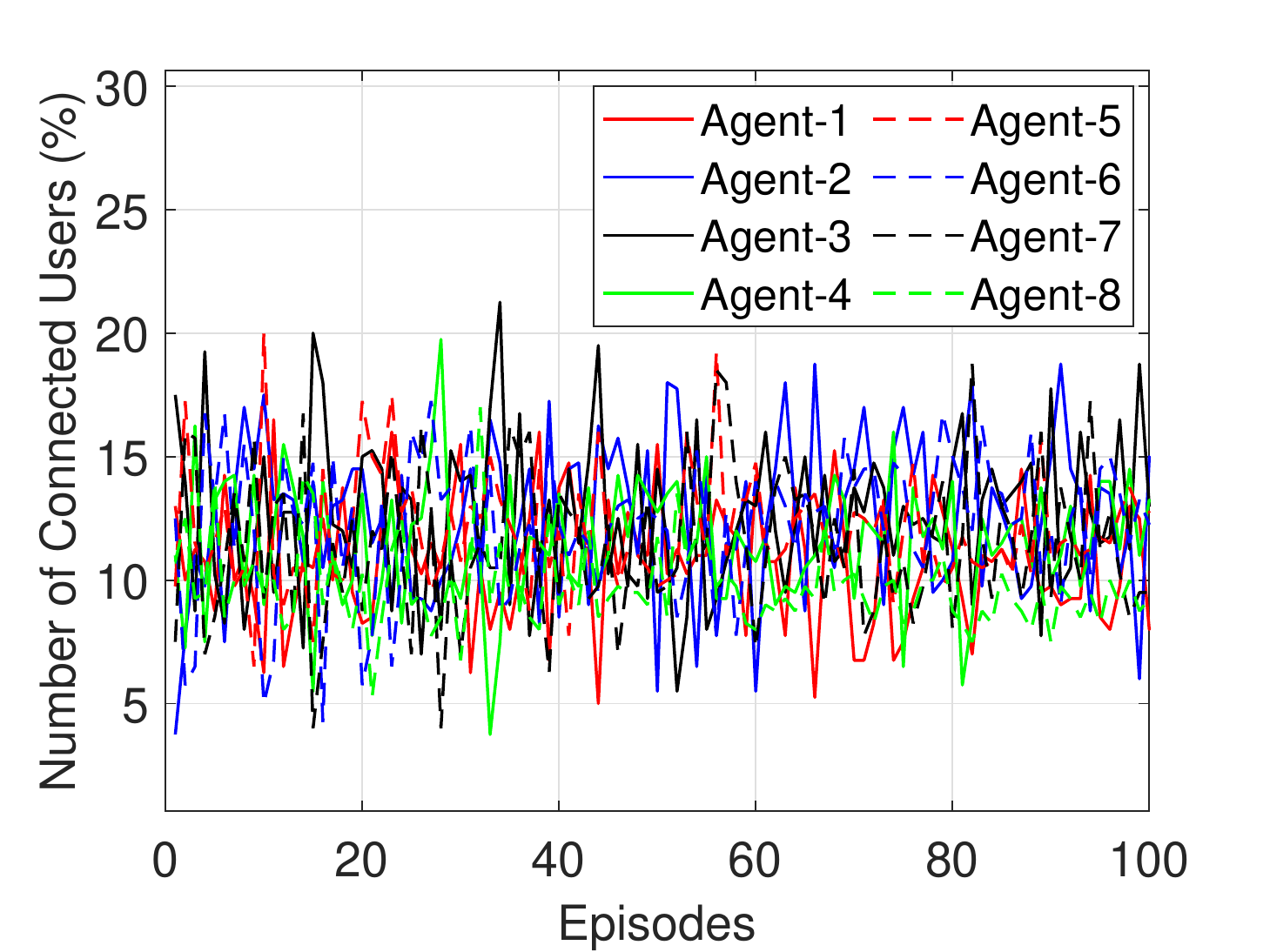}
         \caption{Number of connected users  per agent-controlled UAV vs. episodes.}
         \label{subfig:covVsepi_agents_cmad}
     \end{subfigure}
     \newline
     \begin{subfigure}[b]{0.3\textwidth}
         \centering
         \includegraphics[width=\textwidth]{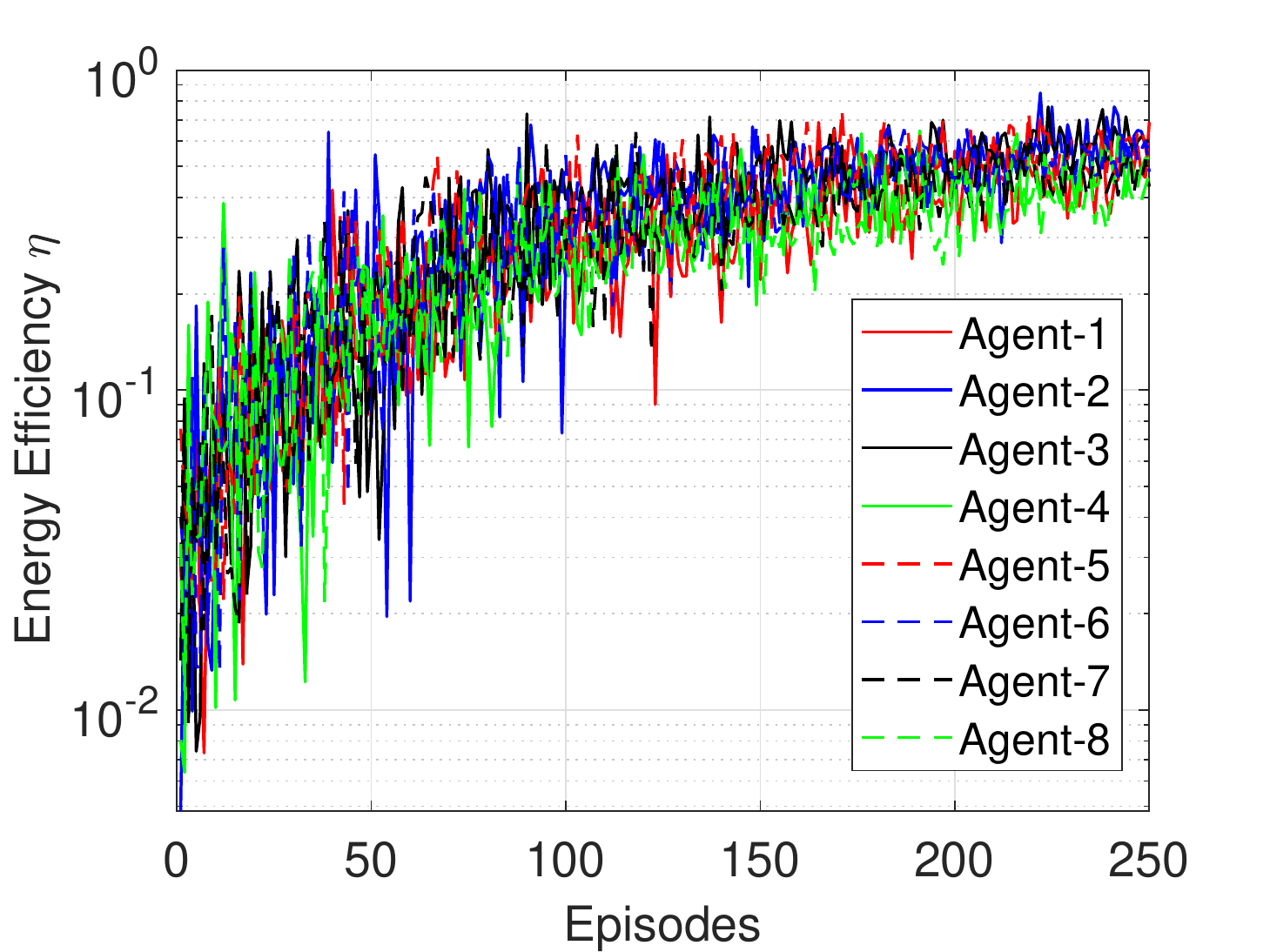}
         \caption{Energy efficiency per agent-controlled UAV vs. episodes.}
         \label{subfig:eeVsepi_cmad}
     \end{subfigure}
     \hfill
     \begin{subfigure}[b]{0.3\textwidth}
         \centering
         \includegraphics[width=\textwidth]{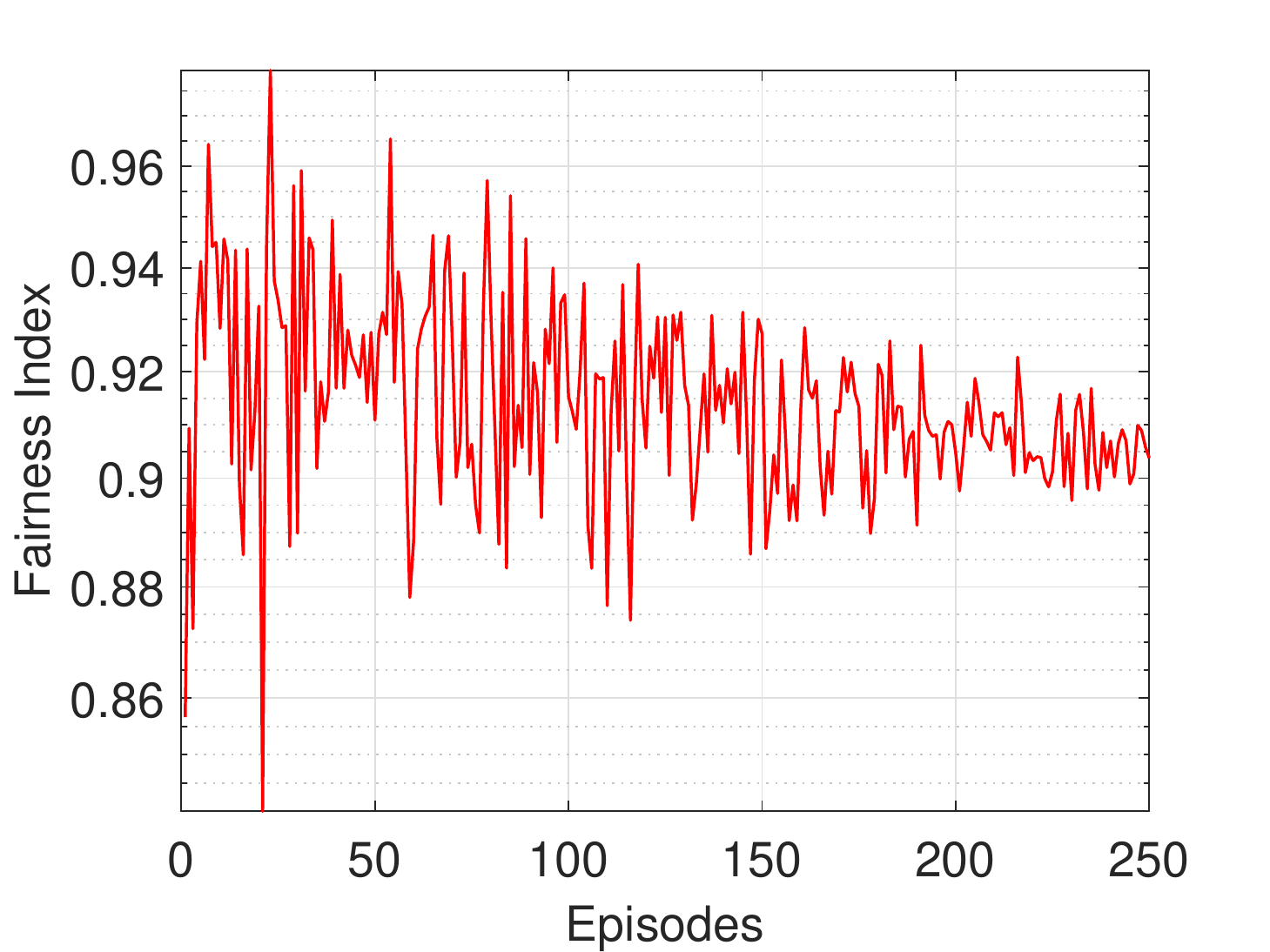}
         \caption{Fairness Index  vs. episodes.}
         \label{subfig:fairVsepi_cmad}
     \end{subfigure}
     \hfill
     \begin{subfigure}[b]{0.3\textwidth}
         \centering
         \includegraphics[width=\textwidth]{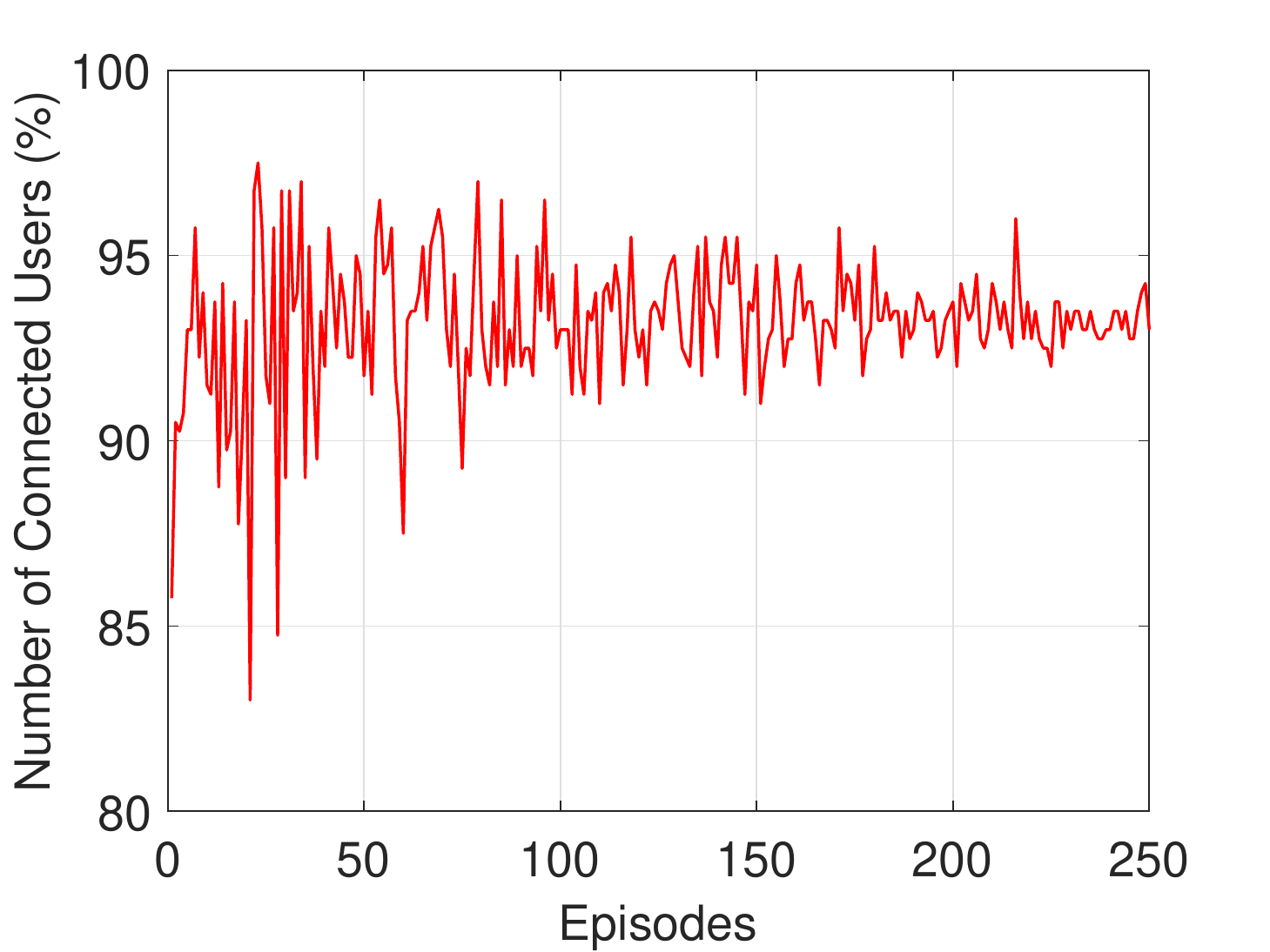}
         \caption{Total number of connected users vs. episodes.}
         \label{subfig:covVsepi_cmad}
     \end{subfigure}
        \caption{Learning behaviour of eight CMAD--DDQN agent-controlled UAVs serving 400 randomly distributed ground users a 1~$km^2$ area of Dublin.}
        \label{fig: Learning_CMAD_DDQN}
\end{figure*}

\subsection{Mobility scenarios}
Due to the difficulty in obtaining non-sparse and temporal mobility traces, we adopt 3 mathematical-based mobility models. These models are extensively used in the ad-hoc networks literature to depict realistic mobility patterns of ground devices~\cite{Song2010, Camp2002}.
\subsubsection{Random Walk mobility model (RW)}
 The RW was developed to mimic the stochastic behaviour of mobile ground devices~\cite{Camp2002}. In RW, the ground devices can change their speed and direction randomly and are uniformly distributed in the range $[0, 2\pi]$ in each time-step $t$ with zero pause time~\cite{Roy2010}. Each movement occurs either in a constant time interval or in a constant travel distance. 
\subsubsection{Random Way-Point mobility model (RWP)}
The RWP is a more realistic mobility model using ad-hoc networks with an introduction of device pause times between changes in direction and/or speed~\cite{Roy2010}. In this model, the travel distance varies in each time-step $t$~\cite{Camp2002}. However, the RW and RWP models were subject to sudden stops, sudden speed changes and sudden changes in the direction where a user can make a sharp 180 degrees turn.
\subsubsection{Gauss–Markov Mobility Model (GMM)}
 The GMM was designed to adapt to different levels of randomness via one tuning parameter~\cite{Camp2002}. The design was inspired by the need for a more realistic mobility model, that is, it allows the users to accelerate, decelerate, or turn progressively while avoiding sharp turns. Initially, each ground user is assigned a current speed and direction. At each time step, movement occurs by updating the speed and direction of each user by following a Gaussian distribution~\cite{Camp2002}.
 
\section{Results}
In this section, we present experimental results based on the settings earlier discussed. We consider the presence of both static and mobile ground devices deployed within the geographical area to depict a realistic dynamic setting and capture the distribution of ground users. Unless stated otherwise, we leveraged location data in the Drumcondra South A area of Dublin with coordinates around 53° 22' 9" N, 6° 14' 45" W~\cite{Dockland_data_bins} along with synthetic data. For the mobility of devices, we adopt different mobility scenarios that depict realistic mobility patterns since there was difficulty in obtaining non-sparse and temporal mobility traces around the Dublin area.
%%%%%%%%%%%New%%%%%%%%%%
%%%%%%%%%%fig for comparing number of agents under metrics
\begin{figure*}
     \centering
     \begin{subfigure}[b]{0.49\textwidth}
         \centering
         \includegraphics[width=\textwidth]{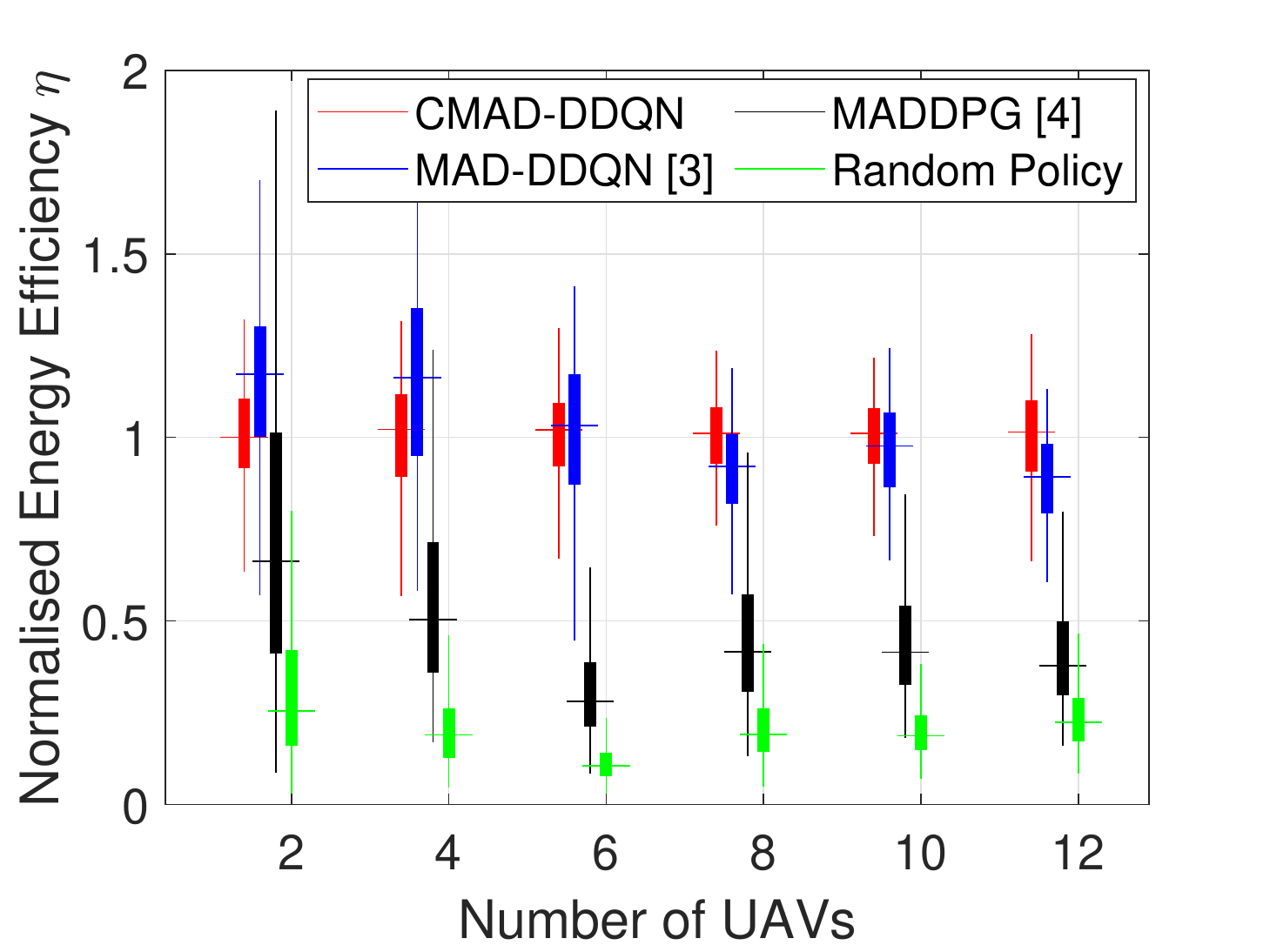}
         \caption{Total Energy efficiency $\eta$ vs. number of UAVs.}
         \label{subfig:EEvsNumUAV}
     \end{subfigure}
     \hfill
     \begin{subfigure}[b]{0.49\textwidth}
         \centering
         \includegraphics[width=\textwidth]{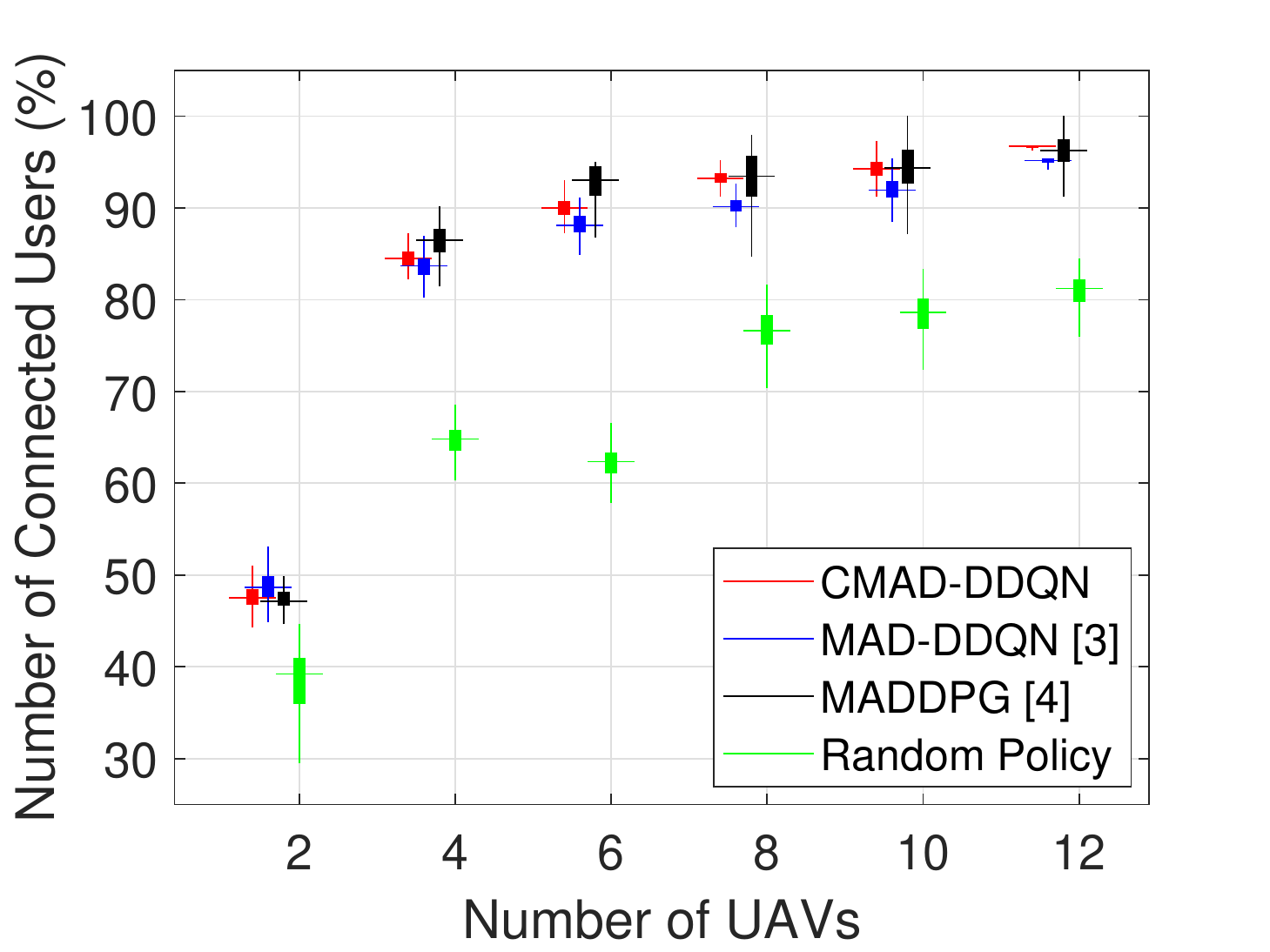}
         \caption{Number of connected ground users vs. number of UAVs.}
         \label{subfig:coveragevsNumUAV}
     \end{subfigure}
     %\hfill
     \newline
     \begin{subfigure}[b]{0.49\textwidth}
         \centering
         \includegraphics[width=\textwidth]{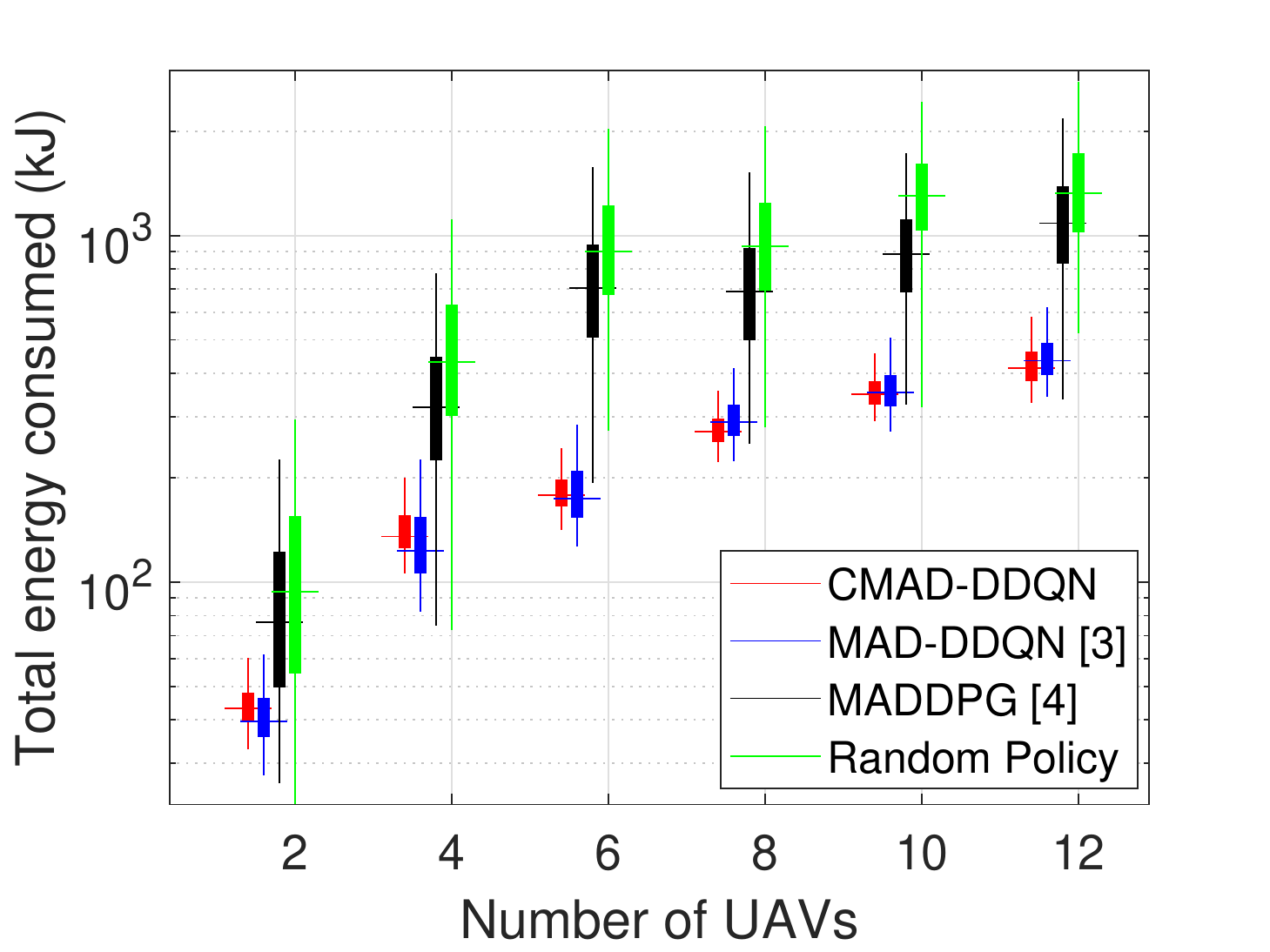}
         \caption{Total energy consumed by UAVs vs. number of UAVs.}
         \label{subfig:energyvsNumUAV}
     \end{subfigure}
     \hfill
     \begin{subfigure}[b]{0.49\textwidth}
         \centering
         \includegraphics[width=\textwidth]{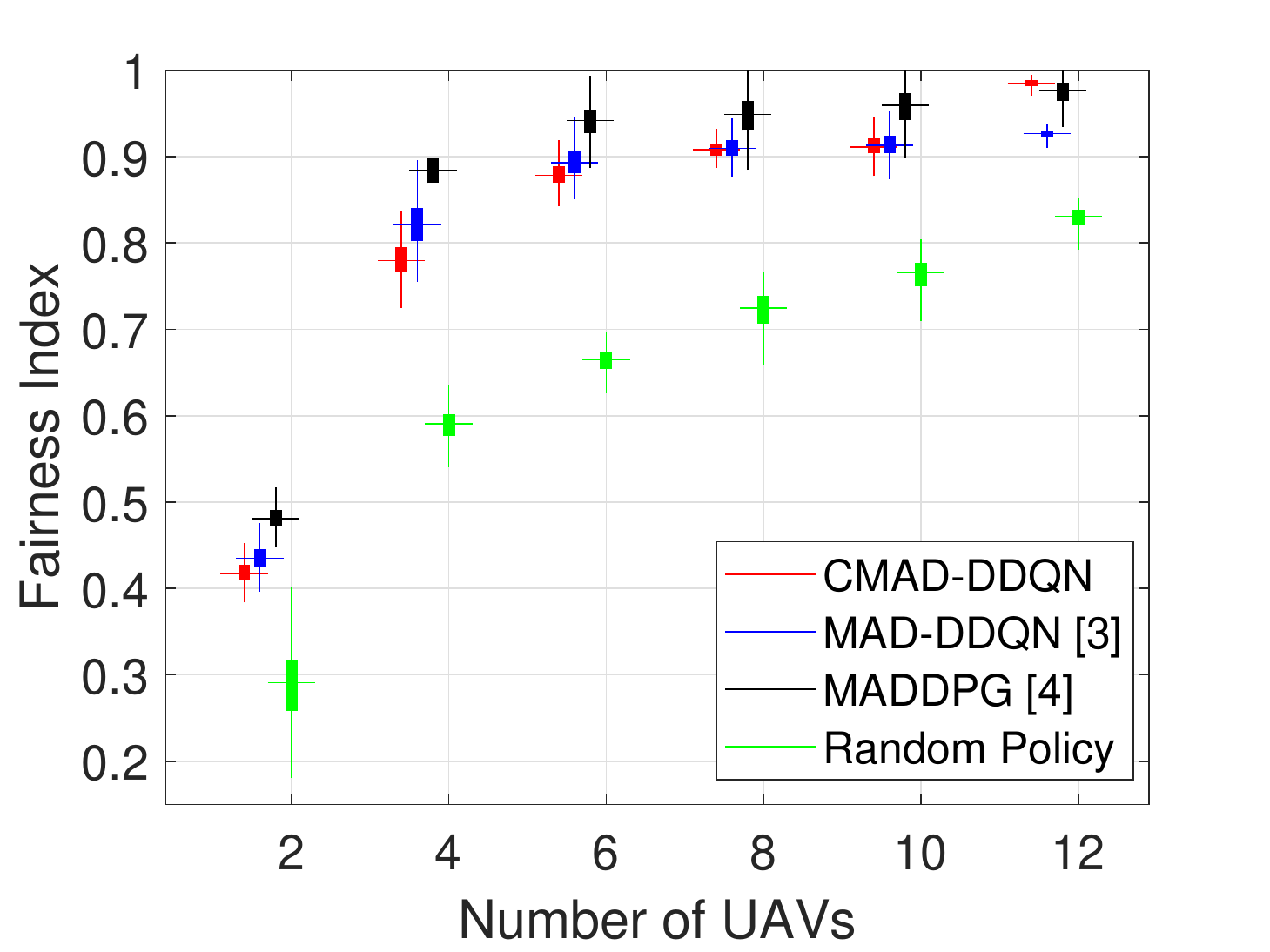}
         \caption{Fairness index vs. number of UAVs.}
         \label{subfig:fairvsNumUAV}
     \end{subfigure}
        \caption{Impact of number of deployed UAVs on the UAVs' EE, number of connected ground users, fairness, and total energy consumed with 200 static and 200 mobile users deployed in a 1 km$^2$ area. The results shown are 2000 runs of trained agents deployed after training.}
        \label{fig: comp_metrics_number}
\end{figure*}

\begin{figure}[h]
    \centering
    \includegraphics[width=3.3in]{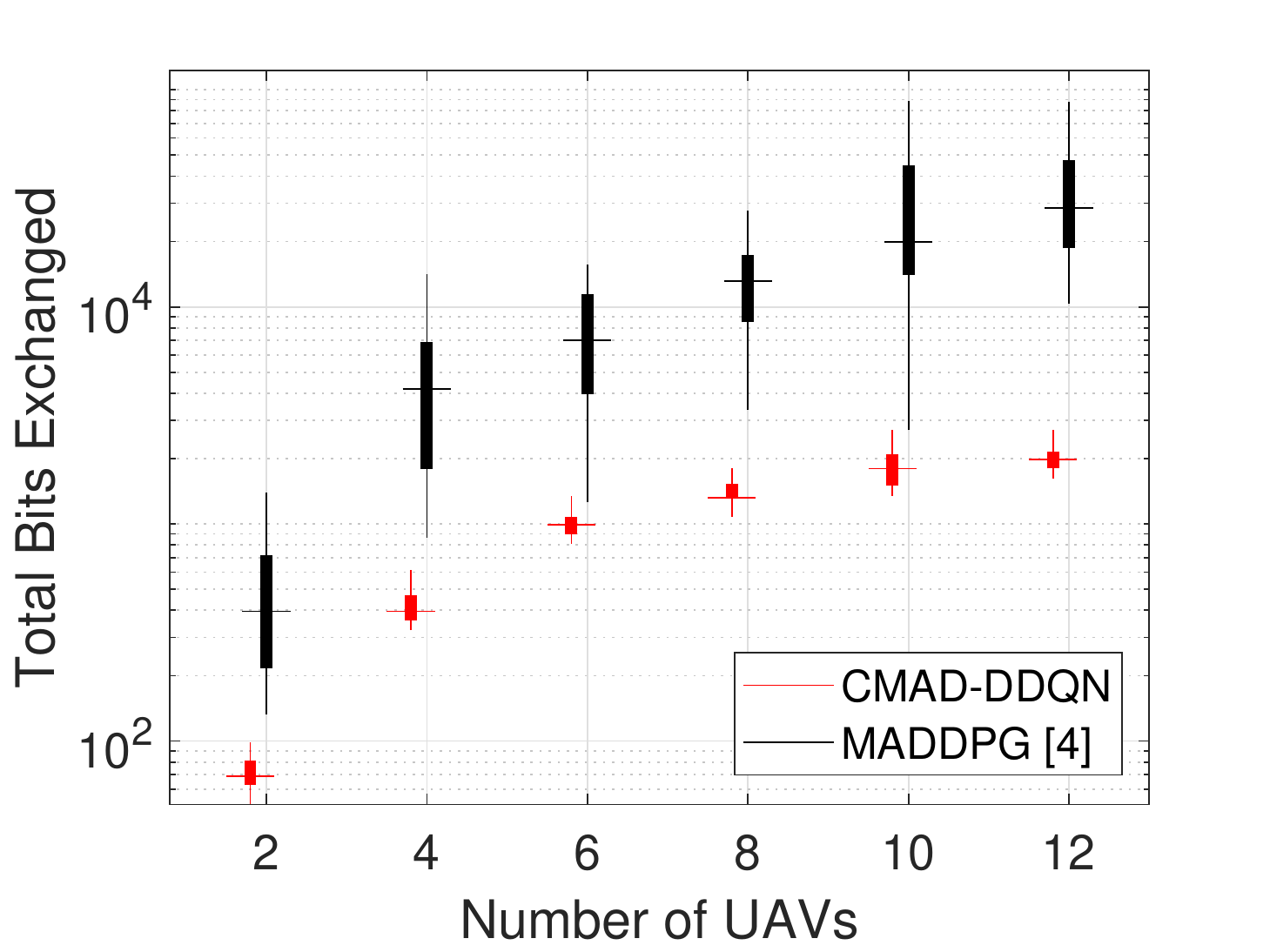}
    \caption{Total bits exchanged vs. number of UAVs. The result evaluates the total overhead incurred by agent-controlled UAVs for decision-making. The results shown are 2000 runs of trained agents deployed after training.}
    \label{fig:overhead}
\end{figure}

%%%%%%%%%%fig for comparing number of agents under metrics
\begin{figure*}
\begin{subfigure}[b]{0.49\textwidth}
\centering
\includegraphics[width=0.9\textwidth]{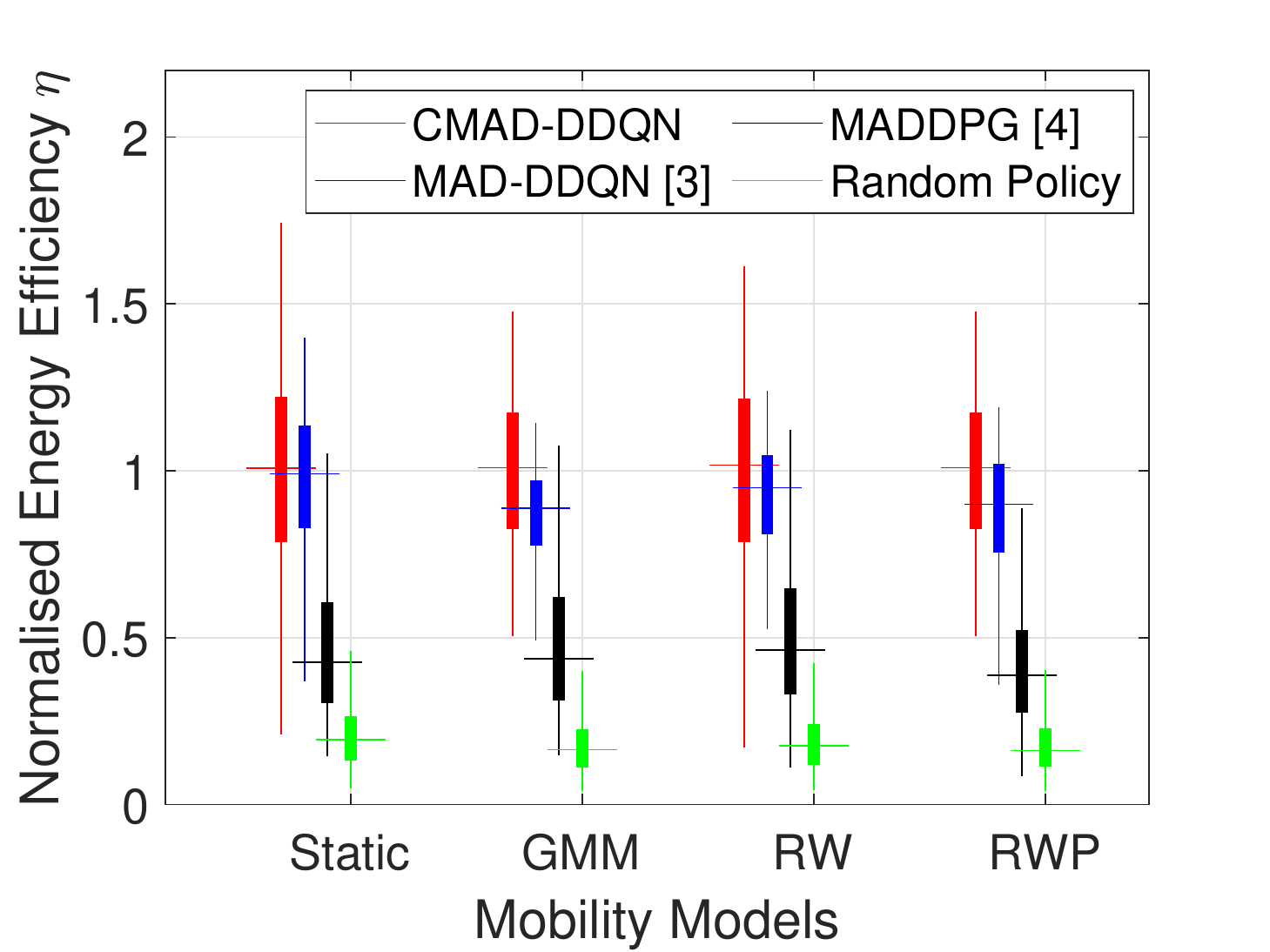}
\caption{Energy efficiency $\eta$ vs. mobility models.}
\label{subfig:EEvsmob}
\end{subfigure}
\hfill
\begin{subfigure}[b]{0.49\textwidth}
\centering
\includegraphics[width=0.9\textwidth]{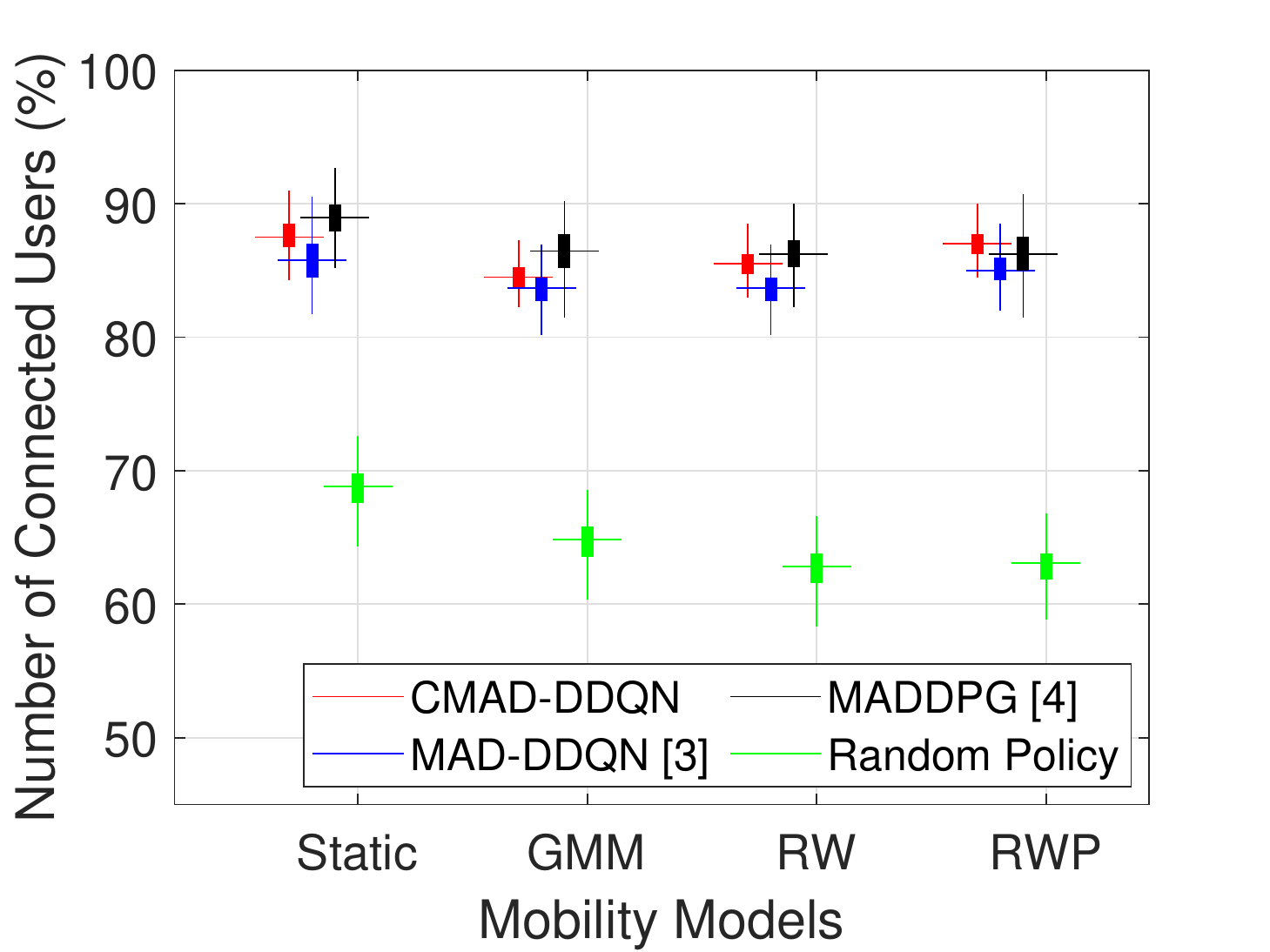}
\caption{Number of connected ground users vs. mobility models.}
\label{subfig:coveragevsmob}
\end{subfigure}
\newline
%\hspace{0.5cm}
\begin{subfigure}[b]{0.49\textwidth}
\centering
\includegraphics[width=0.9\textwidth]{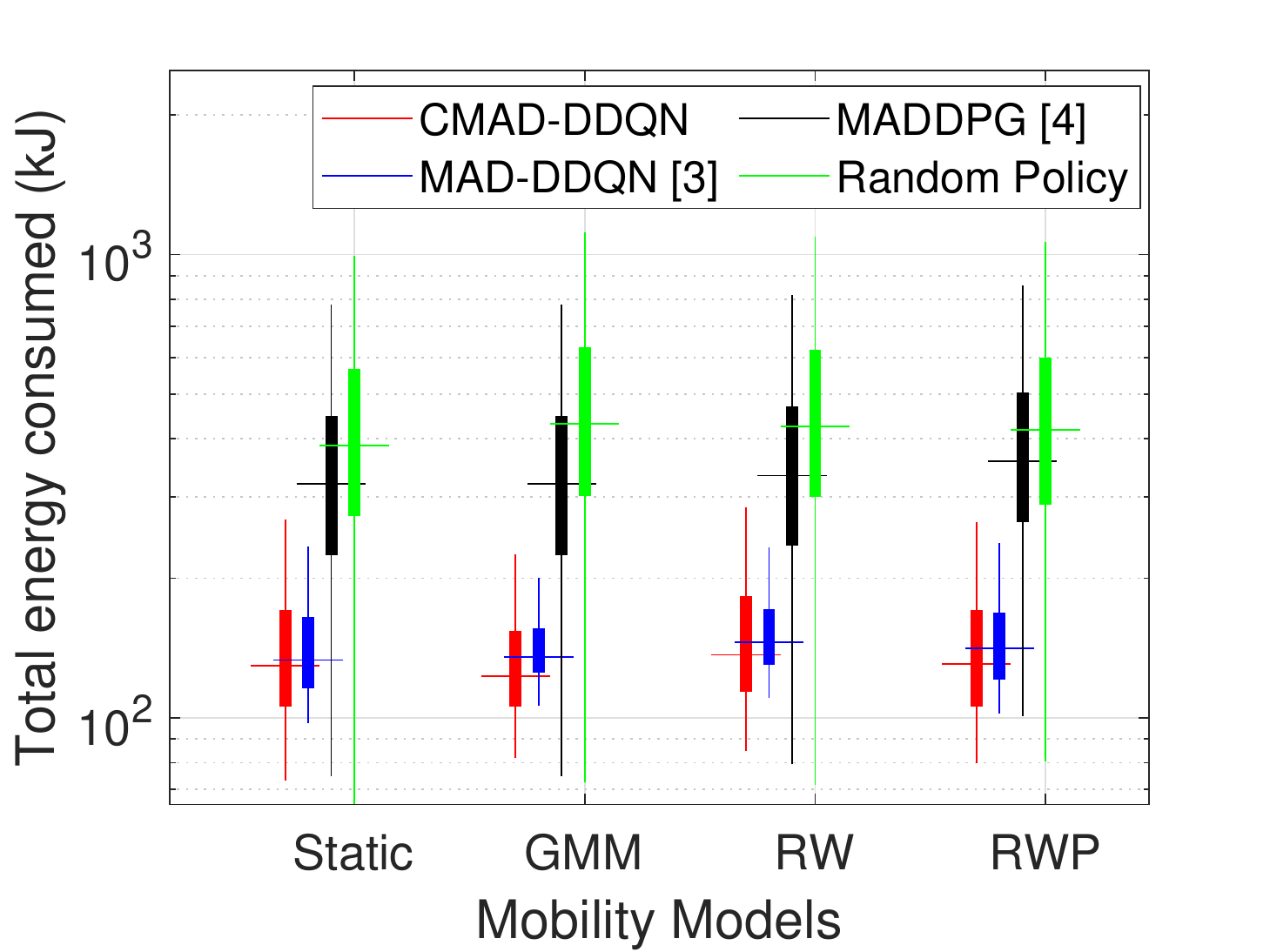}
\caption{Overall energy consumption by UAVs vs. mobility models.}
\label{subfig:energyvsmob}
\end{subfigure}
\hfill
\begin{subfigure}[b]{0.49\textwidth}
\centering
\includegraphics[width=0.9\textwidth]{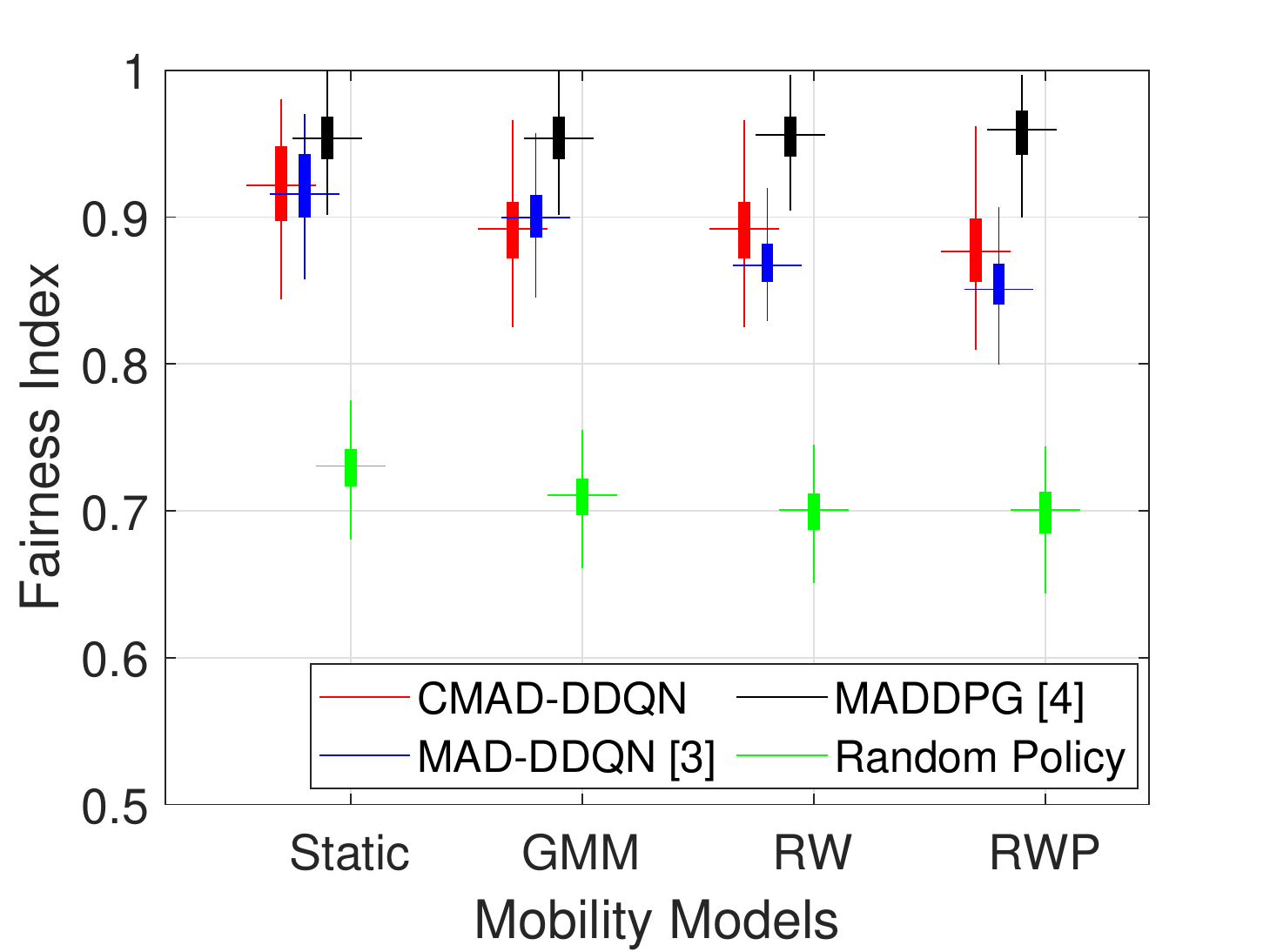}
\caption{Fairness index vs. mobility models.}
\label{subfig:fairvsmob}
\end{subfigure}
\caption{Impact of mobility models of 8 deployed UAVs on the EE, number of connected users, total energy consumed and fairness. For static, we consider 400 static users. For the GMM, RW and RWP we consider 200 static and 200 mobile users following the GMM, RW and RWP mobility models, respectively. The results shown are 2000 runs of trained agents deployed after training.} 
\label{fig: comp_metrics_mobility}
\end{figure*}

%%%%%%%%%%fig for comparing number of agents under metrics
\begin{figure*}%[!htbp]
\begin{subfigure}[b]{0.49\textwidth}
\centering
\includegraphics[width=\textwidth]{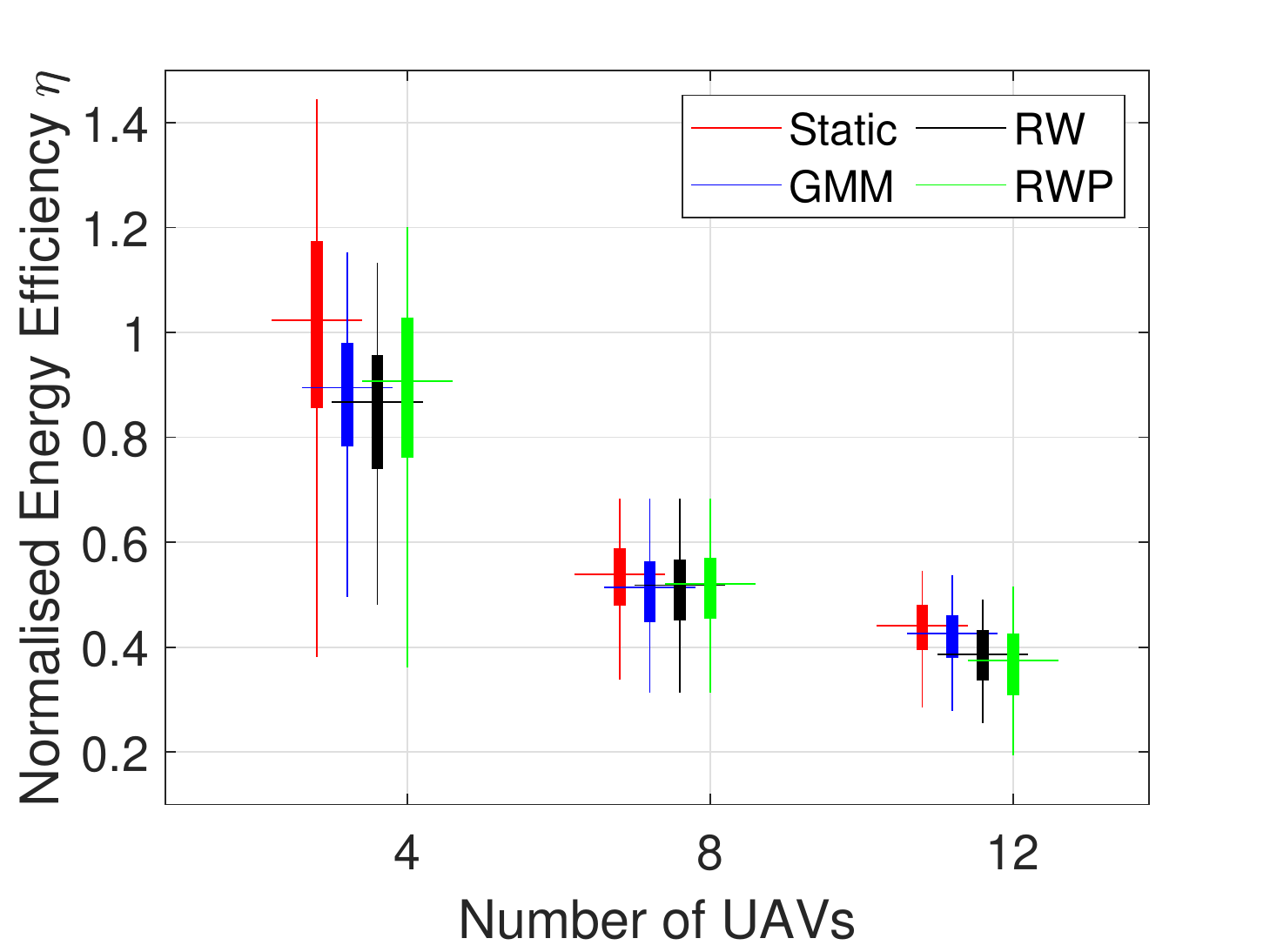}
\caption{Energy efficiency $\eta$ vs. number of UAVs.}
\label{subfig:EEvsNumUAV_mob}
\end{subfigure}
\hfill
\begin{subfigure}[b]{0.49\textwidth}
\centering
\includegraphics[width=\textwidth]{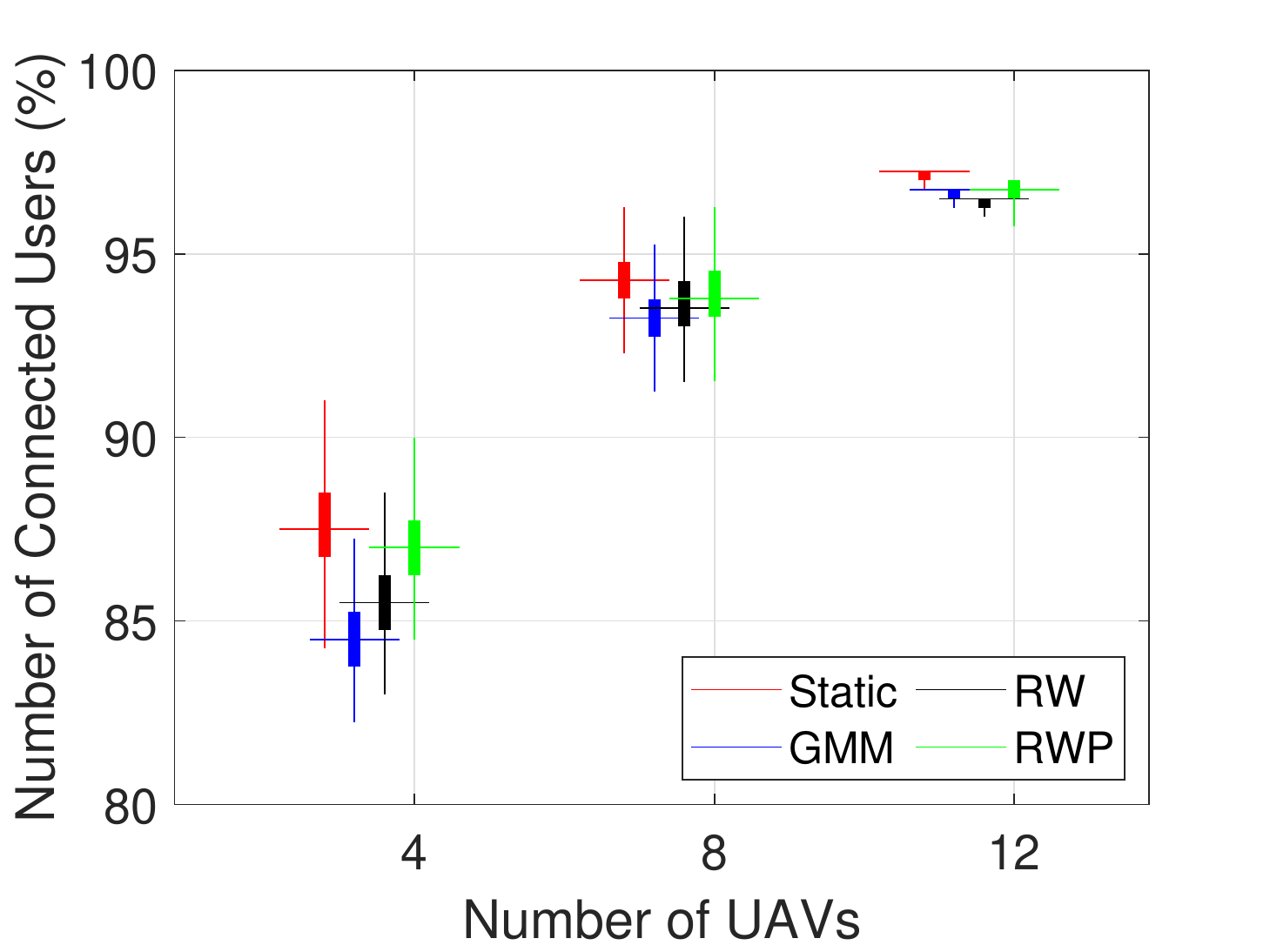}
\caption{Number of connected ground users vs. number of UAVs.}
\label{subfig:coveragevsNumUAV_mob}
\end{subfigure}
\newline
\begin{subfigure}[b]{0.49\textwidth}
%\centering
\includegraphics[width=\textwidth]{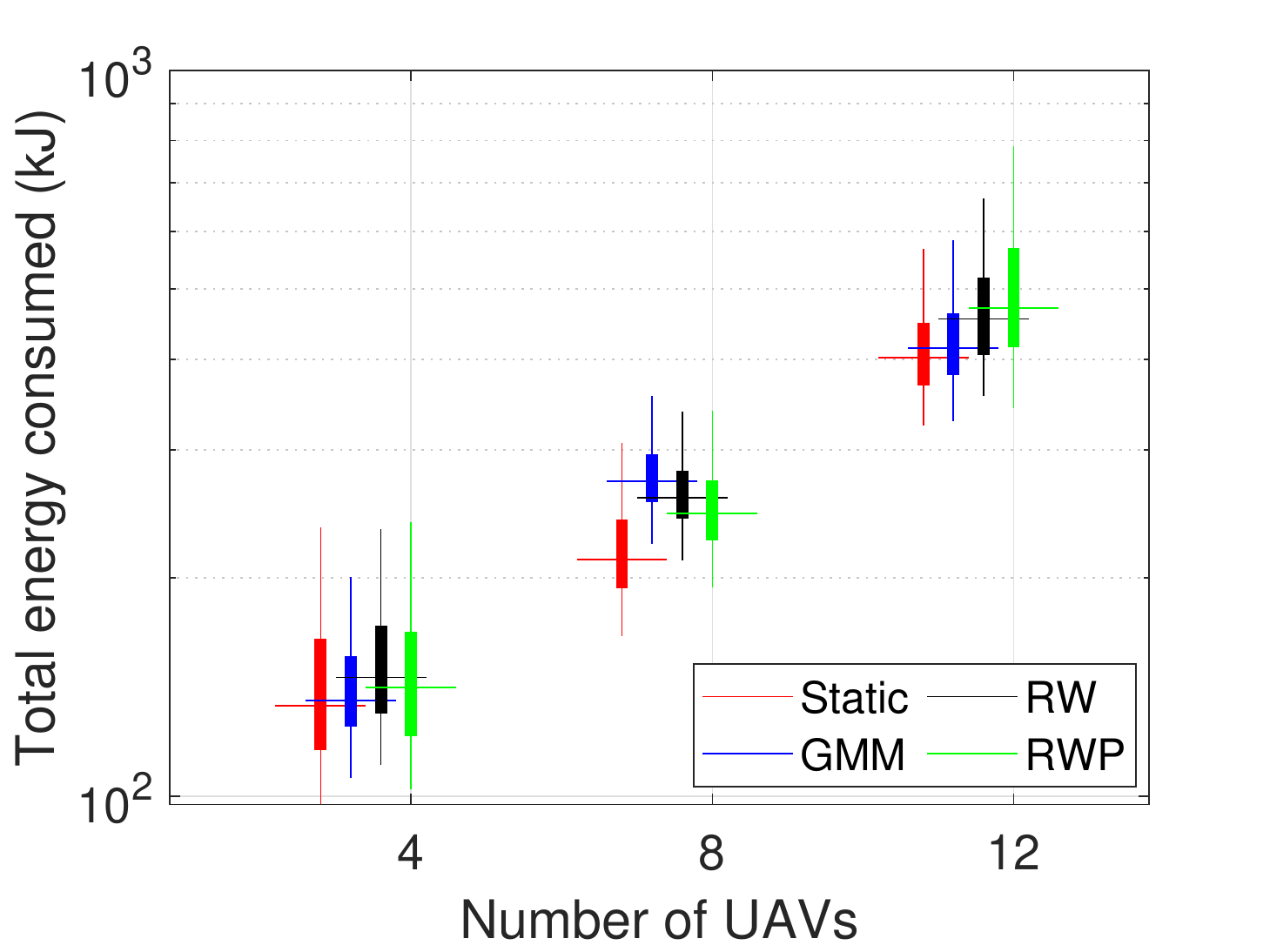}
\caption{Total energy consumed by UAVs vs. number of UAVs.}
\label{subfig:energyvsNumUAV_mob}
\end{subfigure}
\hfill
\begin{subfigure}[b]{0.49\textwidth}
\centering
\includegraphics[width=\textwidth]{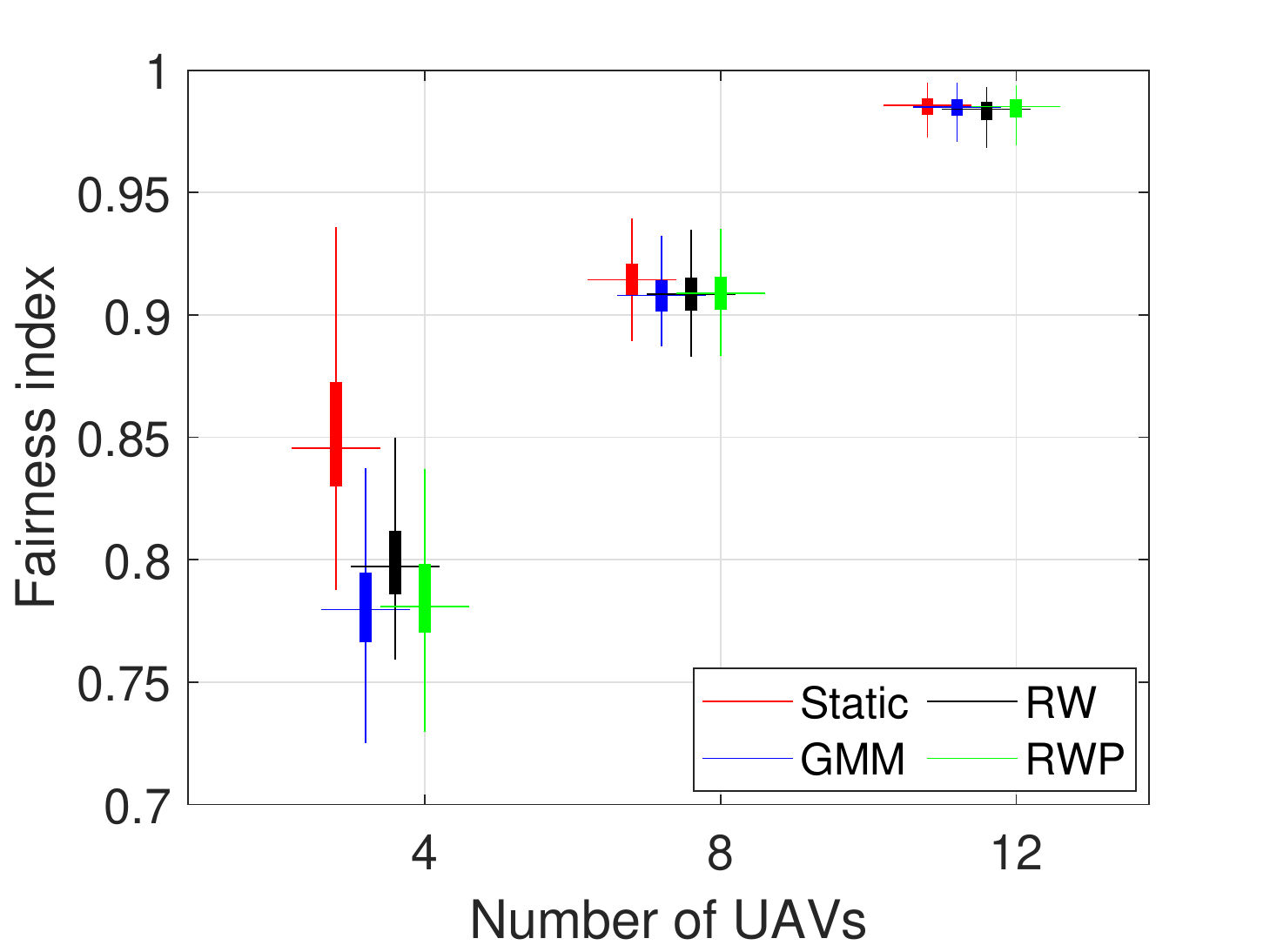}
\caption{Fairness index vs. number of UAVs.}
\label{subfig:fairvsNumUAV_mob}
\end{subfigure}
\caption{Impact of number of deployed UAVs on the UAVs' EE, number of connected ground users, fairness, and total energy consumed while varying mobility scenarios across Drumcondra area of Dublin, Ireland. For static, we consider 400 static users, while for the GMM, RW and RWP we consider 200 static and 200 mobile users following the GMM, RW and RWP mobility models, respectively. The results shown are 2000 runs of trained CMAD--DDQN agents deployed after training.} 
\label{fig: comp_metrics_number_mob}
\end{figure*}

\subsection{Dynamic Setting with Cooperative Agents with Neighbourhood Knowledge}
First, we investigate the learning behaviour of cooperative agents with neighbourhood knowledge as presented in Algorithm \ref{Algorithm_DDQN}. We consider the deployment of eight UAVs serving ground users to observe the performance of the agent-controlled UAVs over a series of time steps. Unlike the MAD--DDQN algorithm, the agents in this variant do have a direct communication mechanism for collaboration. Figure \ref{fig: Learning_CMAD_DDQN} shows the performance of our CMAD--DDQN algorithm measured using the reward, total energy consumed, number of connected users, EE, and fairness index in an environment with randomly deployed static and mobile ground users. As seen in Figure \ref{subfig:RewVsepi_cmad}, all eight agents try to maximise their cumulative reward through the learning episodes. We can see from Figure \ref{subfig:RewVsepi_cmad} that all agents were able to maximise their reward over several time steps. After the $200^{th}$ episode, we observe significant convergence in the energy consumed by the UAVs, within the range of about 16 kJ -- 35 kJ as seen in Figure \ref{subfig:eneVsepi_cmad}. Our proposed system is able to handle UAV failures. In particular, we take into account the possibility of UAVs running out of battery power during deployment. The failure due to the energy outage of a UAV does not in any way impact other UAVs or the training procedure. However, this may only have some impact on the coverage performance, since the operation of more UAVs in the network is expected to provide improved coverage. Figure \ref{subfig:covVsepi_agents_cmad} shows a balance in the connection load across the eight UAVs, ranging between about 6\%--19\% connected ground users per UAV. The total number of connected ground users for all UAVs ranges between 92\%--95\% as seen in Figure \ref{subfig:covVsepi_cmad}. From Figure \ref{subfig:eeVsepi_cmad} and Figure \ref{subfig:fairVsepi_cmad}, we see convergence in the EE and fairness index after the $200^{th}$ episode, respectively.

\subsection{Investigating Number of UAVs Deployment over Baselines}

To observe how the proposed CMAD--DDQN approach performs while deploying varying numbers of UAVs in Figure~\ref{fig: comp_metrics_number}, we compare the proposed CMAD--DDQN approach with baselines to evaluate the impact of different numbers of deployed UAVs on the EE, ground users connectivity and total energy consumed.
Here, we vary the number of UAVs deployed to range between 2 to 12. Since we focus on comparing the EE values rather than showing their absolute values, we normalise the EE values with respect to the mean values of the proposed CMAD--DDQN approach. Figure \ref{subfig:EEvsNumUAV} shows the plot of the normalised EE versus the number of deployed UAVs serving ground users. From Figure \ref{subfig:EEvsNumUAV}, we observe that as more UAVs are deployed, the EE decreases in all approaches possibly because the system becomes more unstable with more UAVs, decreasing the throughput as interference increases, and also takes longer to converge. In particular, the MAD--DDQN approach outperforms the CMAD--DDQN approach in terms of the total system's EE when 2 or 4 UAVs are deployed in the environment. Figure~\ref{subfig:energyvsNumUAV}, we also observe the same trend as the MAD--DDQN approach outperforms the CMAD--DDQN approach in terms of minimising the total energy consumed. However, as more UAVs are deployed, we observe that our proposed CMAD--DDQN approach outperforms the existing baselines in terms of optimising the total system's EE and total energy consumed. Overall, the CMAD--DDQN approach outperforms the MAD--DDQN, MADDPG, and random policy approaches by approximately 15\%, 65\% and~85\%, respectively. Our proposed CMAD--DDQN solution offers improved coordination among the agent-controlled UAVs. From our findings, this improved coordination leads to a corresponding improvement in terms of the overall system's EE, especially as the number of deployed UAVs in the network increased. We observe significant improvement in our proposed CMAD--DDQN approach over the MAD--DDQN approach after deploying 8 UAVs in the network.~However, this improvement in the CMAD--DDQN approach comes with an additional communication overhead as compared to the MAD--DDQN and Random Policy which do not have a direct communication mechanism.~From Figure~\ref{fig: comp_metrics_number}, the communication overhead in the CMAD--DDQN approach results in a slight performance improvement in the evaluation metrics as the number of deployed UAVs is increased.~As expected, the total bits exchanged in the network is increased as the number of agent-controlled UAVs increases as seen in Figure~\ref{fig:overhead}. Furthermore, we observe that our proposed CMAD--DDQN approach performed significantly better than the MADDPG approach in terms of reducing the total amount of bits exchanged during trained deployment. This performance improvement is achieved over the number of UAVs deployments. Intuitively, the overhead incurred by the MADDPG approach is bounded by the overall number of UAVs deployed, thus leading to rapidly-growing control overhead. On the other hand, since our CMAD--DDQN approach considers the communication of an agent-controlled UAV with at most 6 nearest neighbours, we observe a significant reduction in the total amount of bits exchanged after about 8 UAVs are deployed. Nevertheless, we understand that an increase in the control overhead may impact the energy consumption of certain applications.

Figure~\ref{subfig:coveragevsNumUAV} shows the plot of the number of connected users versus the number of deployed UAVs while comparing our proposed CMAD--DDQN approach with the baselines. From Figure~\ref{subfig:coveragevsNumUAV}, we observe a marginally better performance by the MADDPG approach over the CMAD--DDQN and MAD--DDQN approaches in maximising the number of connected ground users by about 0.5\% and 2\%, respectively. However, the slight performance gain by the MADDPG comes at a huge computational training cost which is 8 times higher than the CMAD--DDQN and MAD--DDQN approaches. 
On the other hand, the random policy performed worst among the approaches in reducing connection outages, emphasizing the relevance of strategic decision-making in MARL problems. Figure~\ref{subfig:energyvsNumUAV} illustrates the plot of the total energy consumed versus the number of deployed UAVs serving ground users and clearly shows that the MAD--DDQN and CMAD--DDQN approaches significantly minimise the total energy consumed by all UAVs as compared to the other baselines. Although the MADDPG approach performs better in terms of improving the number of connected users than the random policy, the approach trades energy consumption for improved coverage of ground users. Figure~\ref{subfig:fairvsNumUAV} shows the plot of the geographical fairness versus the number of deployed UAVs serving ground users. The graph shows that fairness improves as the number of UAVs are increased overall approaches. We observed better performance in the fairness index for the MADDPG approach than the CMAD--DDQN and MAD--DDQN approaches when 10 or fewer UAVs are deployed. As 12 UAVs are deployed, the proposed CMAD--DDQN approach outperforms the other baselines in terms of fairness.

\subsection{Investigating Mobility models over Baselines}
In a dynamic scenario where users may be mobile, the UAVs’ locations need to be adjusted in such a way as to improve system performance. In Figure~\ref{fig: comp_metrics_mobility}, we compare the proposed CMAD--DDQN approach with baselines to evaluate the impact of the various mobility models on the EE, number of connected ground users, the geographical fairness and total energy consumed when 8 UAVs are deployed to serve ground users in a 1 km$^2$ area. In Figure~\ref{fig: comp_metrics_number}, we varied the number of deployed UAVs between 2 to 12, however, we chose 8 UAVs as representatives to dig deeper into investigating the impact of different mobility models on the overall system's performance. Figure~\ref{subfig:EEvsmob} shows the plot of the normalised EE versus the mobility models. The ground users' mobility models considered are the Static, GMM, RW and RWP models. Overall deployment of ground users using these mobility models, the proposed CMAD--DDQN approach outperforms the MAD--DDQN, MADDPG and Random Policy approach in terms of maximising the system's EE by about 15\%, 75\% and 85\%, respectively. Figure~\ref{subfig:coveragevsmob} shows how the various mobility models impact the number of connected users while comparing the proposed CMAD--DDQN approach with the baselines. In all mobility models considered, the MADDQN approach performed closely to the proposed CMAD--DDQN approach. However, the MADDQN approach experience very good coverage performance, it had a larger variance than the CMAD--DDQN approach. Our proposed CMAD--DDQN approach converged to a significantly better average over multiple experimental runs. 

Figure~\ref{subfig:energyvsmob} shows the plot of the total energy consumed versus the mobility models while comparing the performance of our proposed CMAD--DDQN with the baselines. Understandably, the random policy consumed the most amount of energy overall mobility models examined. The CMAD--DDQN approach consumes a lesser amount of energy in the static scenario than in the GMM, RW and RWP by about 25\%, 20\% and 15\%, respectively. Although the MADDPG approach performed well in improving the number of connections, it performed poorly in minimizing the total energy consumed. Figure~\ref{subfig:fairvsmob} shows the plot of the geographical fairness versus the mobility models. The CMAD--DDQN approach performed better than the MAD--DDQN approach and random policy but worse than the MADDPG approach. As to be expected, we observed that all approaches performed slightly better in the static scenario, which implies that decision-making in the scenarios that consider the mobility of ground users is worse overall approaches.

\subsection{Investigating the Deployment of UAVs over Mobility Models}

Previously, we see that the mobility of users may have some significant impact on the overall system performance. From Figure~\ref{fig: comp_metrics_mobility} we observe that the proposed CMAD--DDQN approach outperforms other approaches. As we expected, direct communication provides the agents with better insights about the neighbours and may improve the agents' performance. However, communication comes at a cost. Here, we dive deeper to investigate the impact of deploying different numbers of UAVs while varying the mobility model while using CMAD--DDQN algorithm. Figure~\ref{fig: comp_metrics_number_mob} shows graphs of the system's EE, number of connected users, total energy consumed and fairness index versus the number of UAVs while varying the mobility models. Figure~\ref{subfig:coveragevsNumUAV_mob} shows the plot of the number of connected users versus the number of deployed UAVs while varying the mobility models. We observe improve connections as the number of UAVs are increase. Intuitively, more UAV access points can cover more ground users irrespective of the mobility scenario. However, we observe that the static scenario presents us with more connected ground users. We see the plot of the total energy consumed versus the number of deployed UAVs in Figure~\ref{subfig:energyvsNumUAV_mob}, while Figure~\ref{subfig:fairvsNumUAV_mob} shows the plot of the geographical fairness versus the number of deployed UAVs. Intuitively, when more UAVs are deployed to improve coverage, we observe increased geographical fairness, however, this comes at an increased energy cost.

In the case of 4 and 8 UAVs deployment as seen in Figure~\ref{subfig:energyvsNumUAV_mob}, we observe that the RWP model consumes slightly more energy than other models, while the static scenario consumes lesser energy. The fairness index in all mobility scenarios is even up when 12 UAVs are deployed to serve ground users. Figure~\ref{subfig:EEvsNumUAV_mob} shows the plot of the normalised EE versus the number of deployed UAVs. We observe that as more UAVs are deployed, the system's EE drops across all mobility models. The intuition behind that is the increased interference from neighbouring UAVs, that is, as more UAVs are deployed we see an increase in the energy consumed with little increase in the system's throughput.

\section{Conclusion}
In this work, we propose a cooperative communication-enabled multi-agent decentralised double deep Q-network (CMAD--DDQN) approach to optimise the energy efficiency (EE) of a fleet of UAVs serving static and mobile ground users in an interference-limited environment. As we increase the number of UAVs in the network, the system's EE in the CMAD--DDQN approach outperforms existing baselines without degrading the coverage performance, energy utilisation, as well as fairness. The CMAD--DDQN approach guarantees quick adaptability and convergence in a shared and dynamic network environment. The CMAD--DDQN steadily converges faster than the MADDPG approach, thereby leading to better EE. We examine the robustness of the cooperative CMAD--DDQN approach over state-of-the-art approaches while varying various mobility models and observe consistent improvement in the system's EE with a minimally deployed number of UAVs. We demonstrate that the CMAD--DDQN approach significantly outperforms the random policy and state-of-the-art decentralised MARL solutions in terms of EE without degrading coverage performance in the network. Although the periodic exchange of information among agents can dramatically increase the entire system’s communication overheads, it provides a performance guarantee for convergence in most multi-agent systems such as this one. Due to the cost of deploying real-world UAVs, we anticipate that each agent-controlled UAV may be pre-trained and later deployed in real-world environments to carry out coverage tasks. Our future work will investigate if other cooperative methods can offer faster convergence under these dynamic scenarios. We also aim to investigate the impact of different road networks and traffic conditions on the systems' EE.

\section*{Declaration of competing interest}
The authors declare that they have no known competing financial interests or personal relationships that could have appeared to influence the work reported in this paper.

\begin{comment}

\bio{figs/pic1}
Author biography without author photo.
Author biography. Author biography. Author biography.
Author biography. Author biography. Author biography.
Author biography. Author biography. Author biography.
Author biography. Author biography. Author biography.
Author biography. Author biography. Author biography.
Author biography. Author biography. Author biography.
Author biography. Author biography. Author biography.
Author biography. Author biography. Author biography.
Author biography. Author biography. Author biography.
\endbio

\bio{figs/pic1}
Author biography with author photo.
Author biography. Author biography. Author biography.
Author biography. Author biography. Author biography.
Author biography. Author biography. Author biography.
Author biography. Author biography. Author biography.
Author biography. Author biography. Author biography.
Author biography. Author biography. Author biography.
Author biography. Author biography. Author biography.
Author biography. Author biography. Author biography.
Author biography. Author biography. Author biography.
\endbio

\bio{figs/pic1}
Author biography with author photo.
Author biography. Author biography. Author biography.
Author biography. Author biography. Author biography.
Author biography. Author biography. Author biography.
Author biography. Author biography. Author biography.
\endbio
\end{comment}
\end{document}